\documentclass{aa}  
\usepackage{graphicx}
\usepackage{relsize}
\usepackage{txfonts}
\usepackage{xcolor}
\usepackage{fancyvrb}
\usepackage{multirow}   

\definecolor{bblue}{HTML}{002099} 
\definecolor{rred}{HTML}{F22616}  
\definecolor{ggren}{HTML}{00693E} 
\usepackage[breaklinks,colorlinks,citecolor=bblue,linkcolor=bblue,urlcolor=bblue]{hyperref}
\usepackage[capitalise]{cleveref}

\usepackage[all]{hypcap} 

\newcommand\fracAGN{{\tt fracAGN}}
\newcommand\accretion{{\tt agn.accretion\_power}}

\begin{document} 
   \title{A Cigale module tailored (not only) for Low-Luminosity AGN}
        \author{I.\,E.\,L\'opez\inst{\ref{ins:difa},\ref{ins:oas}}\thanks{e-mail: ivan.lopez@inaf.it}
        \and
        G.\,Yang\inst{\ref{ins:niaot},\ref{ins:groningen},\ref{ins:sron}}\thanks{e-mail: gyang@niaot.ac.cn}
        \and
        G.\,Mountrichas\inst{\ref{ins:Cantabria}}
        \and
        M.\,Brusa\inst{\ref{ins:difa},\ref{ins:oas}}
        \and
        D.\,M.\,Alexander\inst{\ref{ins:durham}}
        \and
        R.\,D.\,Baldi\inst{\ref{ins:ira}}
        \and
        E.\,Bertola\inst{\ref{ins:arcetri}}
        \and
        S.\,Bonoli\inst{\ref{ins:dipc},\ref{ins:ikerbasque}}
        \and
        A.\,Comastri\inst{\ref{ins:oas}}
        \and
        F.\,Shankar\inst{\ref{ins:southampton}}
        \and 
        N.\,Acharya\inst{\ref{ins:dipc},\ref{ins:upvehu}}
        \and
        A.\,V.\,Alonso\,Tetilla\inst{\ref{ins:southampton}}
        \and 
        A.\,Lapi\inst{\ref{ins:sissa}}
        \and
        B.\,Laloux\inst{\ref{ins:durham},\ref{ins:athens}}
        \and
        X.\,López\,López\inst{\ref{ins:difa},\ref{ins:oas}}
        \and
        I.\,Muñoz\,Rodríguez\inst{\ref{ins:southampton},\ref{ins:athens}}
        \and
        B.\,Musiimenta\inst{\ref{ins:difa},\ref{ins:oas}}
        \and
        N.\,Osorio\,Clavijo\inst{\ref{ins:IRyA}}
        \and
        L.\,Sala\inst{\ref{ins:unimunich}}
        \and
        D.\,Sengupta\inst{\ref{ins:difa},\ref{ins:oas}}
        }

   \institute{
            Dipartimento di Fisica e Astronomia "Augusto Righi", Università di Bologna, via Gobetti 93/2, 40129 Bologna, Italy \label{ins:difa}
            \and
            INAF - Osservatorio di Astrofisica e Scienza dello Spazio di Bologna, via Gobetti 93/3, 40129 Bologna, Italy \label{ins:oas}
            \and
            Nanjing Institute of Astronomical Optics and Technology, Nanjing 210042, China\label{ins:niaot}
            \and
            Kapteyn Astronomical Institute, University of Groningen, P.O. Box 800, 9700 AV Groningen, The Netherlands\label{ins:groningen}
            \and
            SRON Netherlands Institute for Space Research, Postbus 800, 9700 AV Groningen, The Netherlands\label{ins:sron}
            \and
            Instituto de Física de Cantabria (CSIC-Universidad de Cantabria), Avenida de los Castros, 39005 Santander, Spain\label{ins:Cantabria}
            \and
            Centre for Extragalactic Astronomy, Department of Physics, Durham University, Durham, DH1 3LE, UK\label{ins:durham}
            \and         
            INAF - Istituto di Radioastronomia, via Gobetti 101,  40129 Bologna, Italy\label{ins:ira}
            \and
            INAF - Osservatorio Astrofisco di Arcetri, Largo E. Fermi 5, 50125, Firenze, Italy\label{ins:arcetri}
            \and
            Donostia International Physics Center (DIPC), Manuel Lardizabal Ibilbidea, 4, Donostia-San Sebastián, Spain\label{ins:dipc}
            \and
            IKERBASQUE, Basque Foundation for Science, E-48013, Bilbao, Spain\label{ins:ikerbasque}
            \and
            School of Physics and Astronomy, University of Southampton, Highfield, Southampton SO17 1BJ, UK\label{ins:southampton}
            \and
            SISSA, Via Bonomea 265, 34136 Trieste, Italy \label{ins:sissa}
            \and
            Institute for Astronomy \& Astrophysics, National Observatory of Athens, V. Paulou \& I. Metaxa 11532, Greece\label{ins:athens}
            \and
            Instituto de Radioastronomía y Astrofísica (IRyA-UNAM), 3-72 (Xangari), 8701, Morelia, Mexico\label{ins:IRyA}
            \and
            University of the Basque Country UPV/EHU, Department of Theoretical Physics, Bilbao, E-48080, Spain \label{ins:upvehu}
            \and
            Universitäts-Sternwarte, Fakultät für Physik, Ludwig-Maximilians-Universität, Scheinerstr.1, 81679 München, Germany\label{ins:unimunich}
            }


\abstract{
The spectral energy distribution (SED) of low-luminosity active galactic nuclei (LLAGN) presents unique challenges due to their emissions being comparable to their host galaxy's radiation and the complexity of the accretion physics involved. In this study, we introduce a novel CIGALE module specifically designed to address these challenges. The module combines the empirical $L_\mathrm{X}$--$L_\mathrm{12\mu m}$ relationship with physically motivated accretion models, such as advection-dominated accretion flows (ADAFs) and truncated accretion disks, providing a more accurate depiction of LLAGN's central engine emissions. A mock analysis of the module revealed good recovery of true parameters, with only a slight bias toward higher input values, further validating its reliability.

We tested the module on a sample of 50 X-ray-detected local galaxies, including LINERs and Seyferts, where it demonstrated good estimation of bolometric luminosities, even in the presence of significant galaxy contamination. Notably, the previous X-ray module failed to provide AGN solutions for this sample, stressing the need for a novel approach. Comparisons with mid-luminosity AGN from COSMOS and SDSS datasets confirm the module’s robustness and applicability to AGN up to $L_\mathrm{X}<10^{45}$ erg/s. We also expanded the X-ray to bolometric correction formula, making it applicable to AGN spanning ten orders of magnitude in luminosity, and revealing lower $k_\mathrm{X}$ values for LLAGN than typically assumed.

Additionally, our analysis of the $\alpha_\mathrm{ox}$ index, representing the slope between UV and X-ray emissions, uncovered trends that differ from those observed in high-luminosity AGN. Unlike quasars, where $\alpha_\mathrm{ox}$ correlates with $\lambda_\mathrm{Edd}$, LLAGN exhibit nearly constant or weakly correlated $\alpha_\mathrm{ox}$ values, suggesting a shift in accretion physics and photon production mechanisms in low-luminosity regimes.

These results underscore the importance of a multiwavelength approach in AGN studies and reveal distinct behaviors in LLAGN compared to quasars. Our findings significantly advance the understanding of LLAGN and offer a comprehensive framework for future research aimed at completing the census of the AGN population.}


\maketitle

\section{Introduction}
\label{sec:1}

Active galactic nuclei (AGN) are extremely luminous objects fueled by accretion onto supermassive black holes at the cores of galaxies \citep{AlexanderHickox2012NewAR..56...93A}. Their influence significantly shapes the evolution of host galaxies \citep[e.g.,][]{Fabian12} and plays a crucial role in regulating star formation and feedback processes \citep[e.g.,][]{HeckmanBest14}. AGN exhibit diverse spectral and temporal variability, posing challenges for their classification and comprehension. A standard categorization method is based on luminosity, where typical AGN, like quasars (QSOs), have bolometric luminosities exceeding $10^{44}\,\mathrm{erg\,s^{-1}}$, while low-luminosity AGN (LLAGN) display lower luminosities \citep{Peterson97}.

LLAGN can be defined as galaxies with X-ray luminosities less than $10^{42}\,\mathrm{erg\,s^{-1}}$ and additional evidence of nuclear activity, such as AGN-like spectra \citep{Ptak2001AIPC..599..326P}. This category often includes Low-Ionization Nuclear Emission-line Regions \citep[LINERs,][]{Ho1999AdSpR..23..813H}. Seyferts, with generally higher luminosities, are typically categorized directly as AGN. However, exceptions exist, as seen in the case of low-luminosity Seyferts like NGC~1566 \citep{Aguero2004A&A...414..453A}. Behind these blurred boundaries, the community keeps debating whether LLAGN are downscaled versions of their brighter counterparts \citep[e.g.,][]{Maoz07} or whether they are dominated by different accretion physics \citep[e.g.,][]{Ishibashi2014MNRAS.443.1339I}.

In the local Universe, LLAGN are more common than QSOs. The faint-end slope of the luminosity function of AGN suggests that LLAGN could be more prevalent at all redshifts than previously assumed, as indicated by various studies \citep[e.g.,][]{Schawinski09,Aird12}. Despite their prevalence, their nature is not fully understood due to lower luminosity compared to brighter AGN \citep{Padovani2017,Hickox&Alexander2018ARA&A..56..625H}. While AGN are known to significantly influence galaxy evolution through AGN feedback mechanisms, the potential role of LLAGN in galaxy evolution has not been fully elucidated. 

LLAGN can hide compact jets and release kinetic energy through them \citep[e.g.,][]{FernandezOntiveros23}. Unlike the massive jets observed in radio-loud AGN, the lower-power jets from LLAGN may remain confined within the host galaxy, substantially affecting the interstellar medium (ISM) and altering star formation processes \citep{2018MNRAS.479.5544M}. LINERs can be radio-quiet or radio-loud, with low-accreting black holes that launch sub-relativistic and relativistic jets, respectively \citep{Baldi21LemmingsIII}. Beyond kinetic feedback, winds from sustained radiatively inefficient accretion can also suppress star formation \citep{Almeida2023MNRAS.526..217A}. For instance, in the Sombrero galaxy, winds—driven either by radiatively inefficient accretion or by jets—induce outflows exceeding 1000 km/s \citep{Goold2023arXiv230701252G}. Similarly, in the massive spiral galaxy NGC~4579, a low-power radio jet generates low-velocity shocks and turbulence, significantly influencing the ISM by heating the inner kpc and suppressing star formation \citep{Ogle2024}. 

Although these examples provide insights into the impacts on individual sources, as a collective, LLAGN might play a pivotal role in galaxy evolution, a role that is yet to be fully characterized. The high accretion phase of an AGN lasts only 5-10\% of its duty cycle, with most of its cycle being in a low-accreting phase \citep{Novak2011ApJ...737...26N}. While the luminosity definition of LLAGN allows for the possibility of high-accreting low-mass black holes, the local population of LLAGN generally exhibits low accretion rates \citep[$\log$\,$\lambda_\mathrm{Edd}$\,$<$\,$-3$,][]{Ho2009ApJ...699..626H}. Furthermore, local LLAGN may be the relics of previously high-accreting AGN. 

Standard accretion disks are radiatively very efficient and their emission spans a wide spectrum, with strong UV-optical emission produced by the blackbody radiation of the inner orbits at high temperatures \citep{Shakura1973A&A....24..337S}. Instead, in LLAGN, the accretion process occurs through a radiatively inefficient mode \citep{Xie2012MNRAS.427.1580X}, where gas cannot efficiently radiate its thermal energy due to either low density or strong magnetic fields. This leads to lower radiative efficiency and a harder X-ray spectrum (product of the optically thin free-free emission), distinguishing them from higher-luminosity AGN \citep{Narayan1995ApJ...452..710N}. Additionally, LLAGN lack the big blue bump at the UV band and associated with the blackbody radiation of the inner orbits of a thin accretion disk; instead, they exhibit a red bump at 1-10 microns \citep[][]{Ho08,FernandezOntiveros23}.

At low accretion rates, the accretion flow becomes less dense, optically thin, and less effective at cooling. Consequently, the standard accretion disk (geometrically thin, optically thick disks) becomes unstable and is supplanted by another mechanism. 
Observations and theoretical works point to a change of regime at $\log$\,$\lambda_\mathrm{Edd}$\,$<$\,$-2$ \citep{Narayan1998tbha.conf..148N,Cao2007MNRAS.377..425C}. The exact mechanism driving accretion in LLAGN is still debated, with proposed models including hot, advection-dominated accretion flows (ADAF), adiabatic inflow-outflow solutions (ADIOS), and convection-dominated accretion flows \citep[CDAF; for a review, see][]{YuanNarayan2014ARA&A..52..529Y}. Additionally, standard accretion disks truncated at an inner radius and hybrid models combining truncated disks with hot flows are considered plausible mechanisms \citep{Esin1997ApJ...489..865E,Taam2012ApJ...759...65T,Bu2019ApJ...871..138B}. Clarifying the nature of the accretion flow in LLAGN is essential for understanding their evolution, fueling mechanisms, and the growth of supermassive black holes in the local universe.

The identification and characterization of LLAGN present challenges due to their lower brightness and potential contamination from star formation and X-ray binaries \citep[e.g.,][]{Annuar2020MNRAS.497..229A}. Spectral energy distribution (SED) fitting, utilizing multiwavelength data, emerges as a powerful tool for addressing these challenges. SED fitting can be used to disentangle different physical mechanisms at play from the AGN and its host galaxy, providing insights into accretion disk and dusty torus properties, as well as host galaxy characteristics such as star formation rate (SFR) and stellar mass ($M_\star$). Crucially, SED fitting is the optimal method for obtaining the AGN bolometric luminosities.

Various SED-fitting popular codes are available and have recently been shown to yield similar distributions of general physical parameters \citep{Pacifici23}. One of them is Code Investigating GALaxy Emission \citep[CIGALE\footnote{\url{https://cigale.lam.fr}};][]{Boquien19}. CIGALE leverages the wealth of information in multiwavelength data by incorporating related processes as priors in Bayesian fits. Including the X-ray component in the AGN model, commonly referred to as X-CIGALE and included inside the last CIGALE release, has expanded the capabilities of CIGALE \citep{Yang20,Yang22}. X-CIGALE incorporates a power law to model the intrinsic hot corona emission based on the slope between the X-ray emission at 2~keV and UV at 2500~\AA\ \citep[$\alpha_\mathrm{ox}$, ][]{Tananbaum1979ApJ...234L...9T,Just07}. This relationship is expected due to the accretion disk emission peaking in the UV, with the X-ray emission produced in the corona believed to be Compton up-scattering of the UV photons \citep{Haardt1993ApJ...413..507H}. With this and other improvements, X-CIGALE has been applied to study various aspects of AGN, including star formation history, AGN feedback, and black hole accretion on local mass scaling relations \citep[e.g.,][]{Masoura21,Mountrichas22,LopezIE23}. 
However, in contrast to the $L_\mathrm{2500\AA}$--$\alpha_\mathrm{ox}$ relation well-established for high-luminosity QSOs \citep[e.g.,][]{Lusso10}, the relation for LLAGN remains elusive, perhaps due to intrinsic differences in the accretion process.

A promising avenue for investigating LLAGN is the $L_\mathrm{X}$--$L_\mathrm{12\mu m}$ relation, establishing a connection between intrinsic X-ray luminosity (2-10 keV) and the nuclear 12~micron luminosity, as shown by \citet{Gandhi2009A&A...502..457G}. These emissions share a common bond since mid-infrared (MIR) emission is also reprocessed UV emission. The nuclear dust absorbs the emission generated by accretion and reemits it thermally in the IR. This relation has been systematically explored across various luminosity regimes and seems valid for Seyferts and LINERs in the luminosity regime $L_\mathrm{X}$\,$=$\,$10^{40}$--$10^{45}$~erg~s$^{-1}$ \citep{Asmus15,FernandezOntiveros23}. Additionally, it has already been incorporated into SED fitting methodologies, providing an alternative approach to link X-ray emission with optical-IR emission \citep{Duras20}.

This study introduces a novel X-ray module into the widely utilized SED fitting code CIGALE. This module integrates two principal and independents components: the $L_\mathrm{X}$--$L_\mathrm{12\mu m}$ relationship, employed as a prior, and a new central accretion engine that can vary from an ADAF to a standard accretion disk. \Cref{IRX-sec:2} offers an in-depth examination of the innovative module and validation through a mock analysis. \Cref{IRX-sec:3} demonstrates the application of this module on a X-ray-detected sample of 50 local LLAGN. We elucidate our methodological approach for acquiring X-ray intrinsic fluxes, UV-to-FIR photometry, and SED fitting parameters. Moreover, we present secondary AGN samples to validate the $L_\mathrm{X}$--$L_\mathrm{12\mu m}$ relation. \Cref{IRX-sec:4} validates our methodology comparing the luminosities obtained for all the AGN samples with previous studies. We also analyze the X-ray bolometric correction, extending down to the lower limit of $L_\mathrm{Bol}$\,$=$\,$10^{39}$~erg s$^{-1}$. Furthermore, we address the issue of galaxy contamination and we investigate the $\alpha_\mathrm{ox}$ parameter for LLAGN, noting deviations from QSO extrapolations
. \Cref{IRX-sec:5} synthesizes our findings and conclusions, highlighting how this new CIGALE module furnishes a potent toolkit for probing not only the lower-luminosity regime. This advancement will facilitate the study of supermassive black holes in states of low accretion and is poised to augment future AGN censuses, thereby enriching our comprehension of the role these LLAGN play in the evolution of galaxies.

Throughout this work, we adopt a \citet{2003PASP..115..763C} initial mass function (IMF) and a flat WMAP7 cosmology: $H_\mathrm{0}$\,$=$\,$70.4$~km s$^{-1}$ Mpc$^{-1}$, $\Omega_\mathrm{M}$\,$=$\,$0.27$, and $\Omega_\mathrm{\Lambda}$\,$=$\,$0.73$ \citep{Komatsu2011ApJS..192...18K}.

\section{The Code}
\label{IRX-sec:2}
\subsection{Motivation}
\label{subsec:2.1}

CIGALE is recognized as a powerful multiwavelength SED template fitting tool tailored for extragalactic research \citep{{Boquien19}}. It can accurately derive physical parameters spanning the X-ray to radio spectrum, meticulously accounting for dust attenuation for UV-optical photons from heating sources like stars and AGN activity, and its corresponding infrared re-emission via energy balance. This process involves fitting observed SEDs with a user-generated model library incorporating diverse physical templates. The likelihood distribution is computed using each fit, and a Bayesian-like analysis allows for extracting key physical parameters.

Notable enhancements to CIGALE were done by \citet{Yang20,Yang22}, particularly the integration of an X-ray module and the SKIRTOR AGN model \citep{Stalevski2012MNRAS.420.2756S,Stalevski16}. The X-ray module employs the $\alpha_\mathrm{ox}$--$L_\mathrm{2500\AA}$ relation from \citet{Just07} as a foundational prior, linking UV photons from the accretion disk to X-rays re-produced in the corona. In the standard accretion, the UV-optical photons are inverse-Compton scattered into X-ray energies \citep{HaardtMaraschi1991ApJ...380L..51H}. As a result, a correlation exists between the intrinsic AGN luminosity at $2500\AA$ with $\alpha_\mathrm{ox}$, the SED slope between UV and X-ray bands, defined as:
\begin{equation*}
\alpha_\mathrm{ox} = -0.3838\log\left(\frac{L_\mathrm{2500\AA}}{L_\mathrm{2keV}}\right).
\end{equation*}

The SKIRTOR module encapsulates emissions from three primary AGN components: the accretion disk, polar dust, and a clumpy torus. Notably, for the accretion disk emission—often referred to as the central engine or seed photons—CIGALE can employ two parametric disk models: \citet{Schartmann05} and SKIRTOR\footnote{The SKIRTOR accretion disk is based on the original SKIRTOR model \citep{Stalevski2012MNRAS.420.2756S} and has been improved to better match observational data \citep{Feltre2012MNRAS.426..120F}.}. Both disk models share a similar overall shape and include a free parameter ($\delta_\mathrm{AGN}$), which can be used to adjust the power-law slope within the 0.125 to 10-micron range.


Despite its efficacy across a broad wavelength range and its substantial contributions to galaxy astrophysics, CIGALE's application to LLAGN reveals notable limitations. \citet{Ho08} show fundamental discrepancies in LLAGN SEDs relative to QSOs, notably below $\log~\lambda_\mathrm{Edd}$\,$<$\,$-3$, characterized by a dominant red bump over the expected big blue bump. These differences complicate the direct application of QSO models to LLAGN, as the seed photons are different due to their distinct accretion regimes. Currently, CIGALE can only reproduce AGN with the standard disk, making it unable to replicate the SED observed for LLAGN.

Changes in the overall shape of the SED for AGN and LLAGN also affect the relation that CIGALE uses as prior (ie., $\alpha_\mathrm{ox}$--$L_\mathrm{2500\AA}$). While \citet{Maoz07} observed similarities between LLAGN and QSOs, suggesting potential model applicability, divergences in accretion flow-dominated UV photon populations in LLAGN challenge the universal application of the $\alpha_\mathrm{ox}$--$L_\mathrm{2500\AA}$ relation. Additional studies, such as \citet{Xu11}, have explored the $\alpha_\mathrm{ox}$--$L_\mathrm{2500\AA}$ relation for samples of Seyferts and LINERs, revealing certain similarities with established QSO calibrations. However, their determination of $L_\mathrm{2500\AA}$ is through proxies derived from H$\beta$ and B-band ($\sim$4420\,\AA) extrapolation. Even in models assuming LLAGN as scaled versions of QSOs, pure ADAF models deviate slightly from the $L_\mathrm{2500\AA}$--$\alpha_\mathrm{ox}$ relation observed in QSOs \citep{Maoz07}. Furthermore, \citet{EsparzaArredondo20} demonstrated that certain faint AGN deviate from the expected relation. \citet{Nemmen14}, through a physically motivated SED fitting, predicted an $\alpha_\mathrm{ox}$ for LLAGN far from extrapolations of its high-luminosity counterparts, emphasizing the influence of radiatively inefficient accretion processes. The complex accretion physics of LLAGN continues to fuel debate, highlighting the need for tailored approaches rather than relying on universally applied relations.



To overcome the challenges posed by the unique accretion physics of LLAGN, we propose a novel X-ray module tailored specifically for these sources, yet usable for a wide variety of mid-luminosity AGN. Recognizing the current uncertainties in their accretion physics and the need for a more customized approach, we propose two main changes: the seed photons from the central engine and the prior that links the X-ray emission to the rest of the SED. 

Nonetheless, the current adaptation of this module introduces a critical caveat. Given that an ADAF has a considerable synchrotron emission it can also be observed in the radio and submillimeter (sub-mm) wavelengths, so careful consideration is required for another important component of the AGN framework: radio jets. Typically, radio jets are accounted for exclusively within radio frequencies for high-luminosity AGN. The radio module extends to the far-IR in CIGALE, modeling the synchrotron emission as a power law. This specificity enables the distinct determination of the spectrum slope, differentiating it from other sources, such as star-forming regions, modeled with another power law with a different slope.

However, the emission from radio jets spans the entire electromagnetic spectrum, not just the radio domain. This omnipresence can complicate SED fittings that omit their modeling, leading to potential contamination of SEDs intended to map the photon production by SMBH accretion processes. As highlighted by \citet{Nemmen14}, while the typical ADAF exhibits a SED characterized by distinct bumps, the average jet emission in LLAGN manifests as a softer, more constant output across the spectrum from 100 microns to X-ray wavelengths. This spectral behavior makes it impossible to fit without a constraint from the radio bands. Because of that, in this work, jet emission will be neglected in the X-ray to IR analysis.  In future studies, we will incorporate the ADAF emission into the radio module of CIGALE, facilitating the differentiation between potential jet models and ADAFs. This development will allow for precisely fitting and identifying sub-mm and radio emissions in LLAGN, thereby excluding potential synchrotron interference in the broader spectrum. Such advancements will enable us to avoid possible synchrotron emission in the rest of the spectra and thus model the total emission from an AGN more accurately and better understand the role of jets in LLAGN activity.

\subsection{The new module}
\label{subsec:2.2}

We have introduced two main modifications to enhance the utility of CIGALE's AGN model for LLAGN. These changes are designed to provide a more accurate representation of the physical processes occurring in these systems, improving both the theoretical framework and its practical application. 

\begin{figure}
    \centering
    \includegraphics[width=1\linewidth]{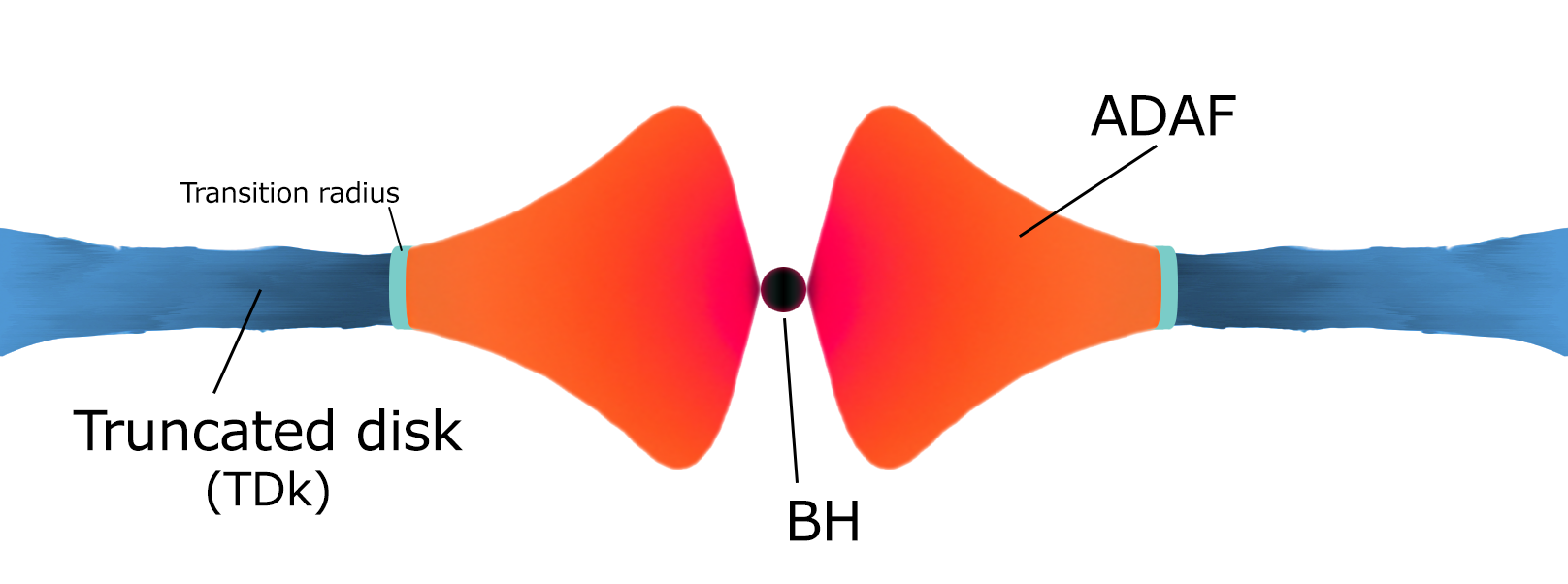}
    \caption[ADAF diagram]{Schematic diagram of an advection-dominated accretion flow (ADAF), the truncated accretion disk (TDk), and the transition radius.}
    \label{fig:adaf_diagram}
\end{figure}

Firstly, the modeling of seed photons emanating from the central engine has been refined by incorporating a combination of a truncated thin disk and an ADAF. A schematic view of the central engine is shown in \cref{fig:adaf_diagram}. Traditional accretion disks, characterized as geometrically thin and optically thick, generate a SED that is the sum of blackbody emission\footnote{The disk can be modeled as a sum of rings, each emitting as a blackbody where its temperature depends on its distance from the SMBH.} \citep{Shakura1973A&A....24..337S}. In contrast, ADAFs are hot, optically thin solutions, and quasi-spherical, with a SED produced by synchrotron radiation, Bremsstrahlung radiation, and inverse Compton scattering. A truncated disk emulates the blackbody spectrum like a standard disk, albeit without the innermost and hotter orbits responsible for UV photon production. The details of the implementation are described in \cref{subsec:2.2.1}.

Secondly, we revisit the connection between the X-ray module and the UV-to-IR SKIRTOR model. The traditional $\alpha_\mathrm{ox}$--$L_\mathrm{2500\AA}$ relation is replaced with an empirical $L_\mathrm{X}$--$L_\mathrm{12\mu m}$ relation \citep[][]{Asmus15}. This relation connects the intrinsic X-ray luminosity (2-10 keV) and the nuclear 12~micron luminosity. X-ray emission from the host galaxy is implemented in the same way as \citet{Yang22}. The details of this implementation are outlined in \cref{subsec:2.2.2}.

\subsubsection{A new central engine}
\label{subsec:2.2.1}

For the pure-ADAF spectrum, we parameterize the mean SED based on the template from \citet{Nemmen14}. This module employs a physically motivated code, considering a hot, geometrically thick, optically thin two-temperature accretion flow with primary radiative processes being synchrotron emission, bremsstrahlung, and inverse Compton scattering. Our parameterization is limited to the UV-to-IR bands, as different CIGALE modules address radio and X-ray. 



\begin{figure}[!h]
    \centering
    \includegraphics[width=1\linewidth]{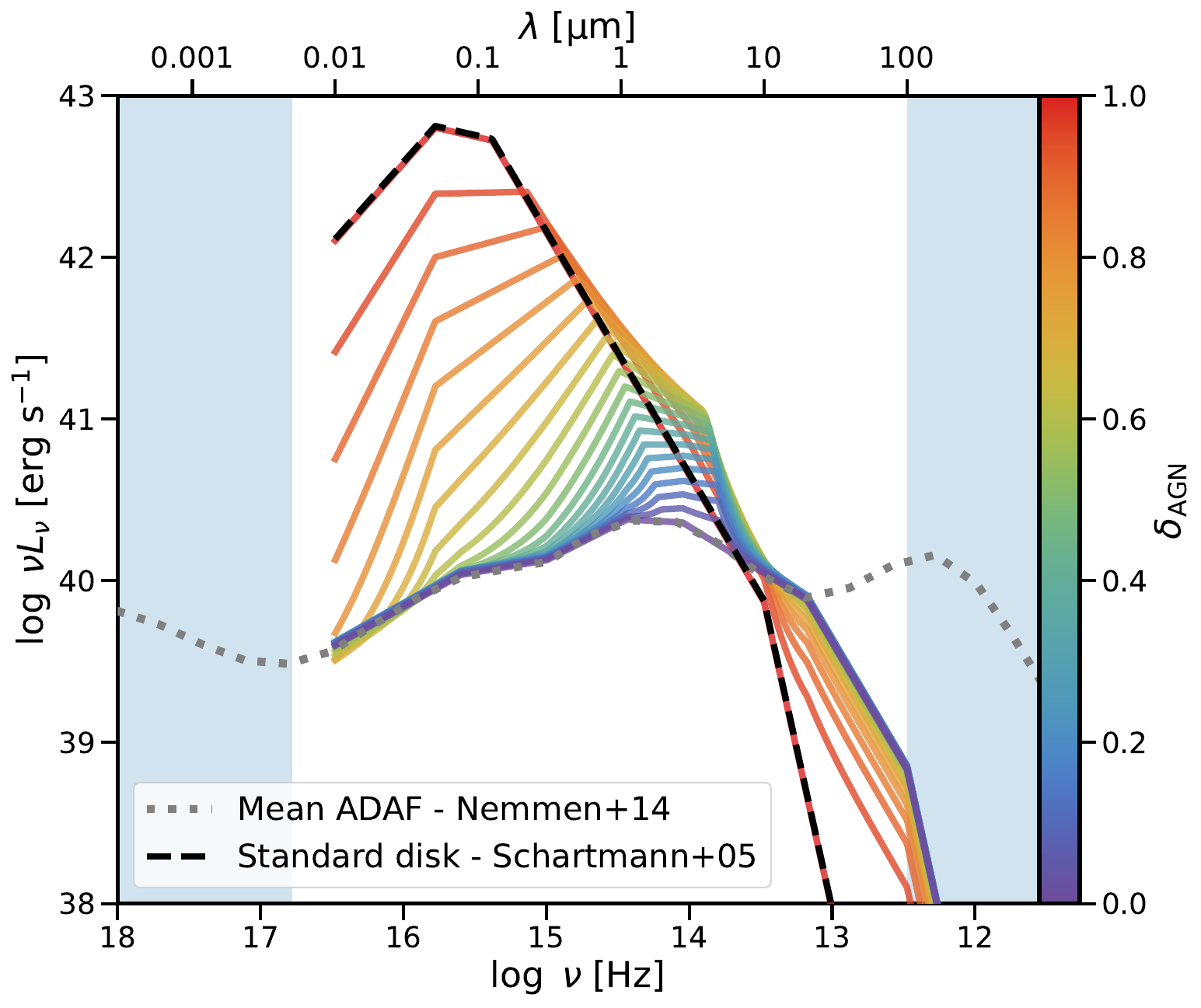}
    \caption{Intrinsec SEDs for the central engine. The dotted green line represents the mean ADAF from \citet{Nemmen14}. The dashed black line depicts the parametrization by \citet{Schartmann05}, commonly used in CIGALE for QSOs with standard accretion disks. The solid lines represent our new parametrization, transitioning from a blue line illustrating the pure ADAF ($\delta_{\mathrm{AGN}}$\,$=$\,$0$)  to a red line indicating a pure standard accretion disk ($\delta_{\mathrm{AGN}}$\,$=$\,$1$). Intermediate values represent a transition where the ADAF and a truncated accretion disk contribute to the SED. At lower $\delta_{\mathrm{AGN}}$, the truncated accretion disk exhibits strong infrared emission, consistent with the cooler outer orbits of an accretion disk. As $\delta_{\mathrm{AGN}}$  increases, the spectrum also includes contributions from the hotter internal orbits, producing more UV-optical photons. The blue-shaded regions denote the wavelength ranges where the other modules of CIGALE operate, specifically X-ray on the left and Radio on the right.}
    \label{fig:ADAF-sed-model}
\end{figure}

\citet{Nemmen14} also explore the possibility of a truncated accretion disk located just outside the ADAF region. These truncated disks are essential for modeling the red bump, commonly observed between 1 and 10 microns, which is absent from the non-thermal continuum \citep[see, for example, the Sombrero galaxy in][]{FernandezOntiveros23}. However, these two components are interdependent, as truncated disks share boundary conditions with the ADAF. Moreover, an external disk reprocesses the X-ray radiation generated by the ADAF. Fortunately, \citet{Nemmen14} demonstrates that this effect is negligible, resulting in a SED identical to the standard disk. Based on this, we separate the two SED components and model the seed photon contribution as follows:

\begin{equation}
\centering
\lambda L_\lambda^\mathrm{Total} \propto \delta_\mathrm{AGN} \times \lambda L_\lambda^\mathrm{ADAF} + (1-\delta_\mathrm{AGN}) \times \lambda L_\lambda^\mathrm{Disk}
\end{equation}

where $\delta_\mathrm{AGN}$ is a free parameter ranging from 0 to 1. When $\delta_\mathrm{AGN} = 0$, the SED represents a pure ADAF contribution, based on the average ADAF model from \citet{Nemmen14}. When $\delta_\mathrm{AGN} = 1$, the SED follows the standard accretion disk model from \citet{Schartmann05}. \Cref{fig:ADAF-sed-model} presents the intrinsic SEDs obtained for varying $\delta_\mathrm{AGN}$ values between 0 and 1.

Intermediate values of $\delta_\mathrm{AGN}$ represent a transition phase, where both the ADAF and the truncated accretion disk contribute to the SED. For $\delta_\mathrm{AGN}$ values between 0 and 0.5, the emission resembles the thermal radiation from the larger orbits of the accretion disk, resulting in a red bump. From 0.5 to 1, the spectrum gradually shifts from an ADAF + truncated disk to the Schartmann parametrization, removing the ADAF contribution while incorporating more inner orbits of the accretion disk. This shift produces an increasing number of UV-optical photons as the contribution from the hotter inner orbits becomes more prominent.

This adaptable model allows independent management of both components, enabling the recreation of a wide range of SEDs, including both regular AGN and LLAGN, and accounts for scenarios with or without a red bump. For mid-luminosity AGN or sources with $\log \lambda_\mathrm{Edd}>-2$, where a standard accretion disk is expected, the $\delta_\mathrm{AGN}$ parameter can be fixed at 1.

\subsubsection{A new way to connect X-ray to the SKIRTOR model}
\label{subsec:2.2.2}

The decision to move away from the $\alpha_\mathrm{ox}$--$L_\mathrm{2500\AA}$ relation is motivated by the lack of reliable calibrations for this relation in the low-luminosity regime. Given that this relation links photons from the accretion disk with reprocessed emissions in the X-ray spectrum, changes in the UV photon population could significantly alter its calibration.

In contrast, the $L_\mathrm{X}$--$L_\mathrm{12\mu m}$ relation is based on a direct linkage between two different reprocessed photon types: X-ray and MIR. This relation has proven valid for a wide variety of AGN types, including Seyferts I, II, and LINERs, across five orders of magnitude up to $L_\mathrm{X}$\,$\sim$\,$10^{45}$~erg~s$^{-1}$. Notably, this relation has been successfully utilized by \citet{Duras20} for SED fitting, providing estimates of bolometric luminosity and corrections across various bands widely accepted in the field.

The $L_\mathrm{X}$--$L_\mathrm{12\mu m}$ relation is founded on the principle that UV-optical photons, originating from accretion physics, are reprocessed into X-rays by the corona and into MIR by nuclear and polar dust \citep{Gandhi2009A&A...502..457G}. Interestingly, this relation appears unaffected by obscuration produced at the torus or the polar dust, exhibiting less than 0.1 dex differences between unobscured and highly obscured objects \citep{Asmus15,Lopez2016A&A...591A..47L,Asmus2016ApJ...822..109A}. This stability is attributed to emissions from the polar outflow regions in the MIR and potentially higher covering factors in more obscured objects, which mitigate differences due to anisotropic viewing angles. This characteristic makes the relation particularly suitable for LLAGN, which, at low Eddington ratios, are believed to have high covering factors \citep[$\sim$\,$80$--$85\%$][]{RamosAlmeidaRicci2017NatAs...1..679R}. 

\textit{Is this relation valid in low-luminosity regimes?} One could argue that although the central engine may change, the reprocessing of UV-optical photons into X-rays and MIR remains similar to that in high-luminosity cases. For instance, \citet{Chakraborty2023A&A...676L..13C} found an anti-correlation between the optical depth and electron temperature of the corona in LLAGN, akin to what is observed in Seyferts with higher accretion rates, suggesting comparable corona physics across AGN of varying luminosities. Furthermore, \citet{Mason2013ApJ...777..164M} demonstrated that nuclear emission at 12 microns in LLAGN primarily originates from dust re-emission heated by UV photons. \citet{Mason2012AJ....144...11M} also showed the applicability of this relation to LLAGN using Spitzer data, while \citet{FernandezOntiveros23} confirmed that LLAGN exhibit minimal scatter from the \citet{Asmus15} relation, based on multiwavelength sub-parsec resolution observations.

An important caveat involves sources without central dust or a torus, as it remains uncertain whether the torus should disappear at lower accretion rates \citep[see][]{Elitzur2008NewAR..52..274E}. However, high-resolution ALMA observations have shown that a pcs-scale dust torus structure still exists in a small sample of LLAGN \citep{Combes2019A&A...623A..79C}. Even in the most extreme cases where dust is absent, LLAGN tend to follow the MIR-X-ray correlation. This alignment is likely due to energy balance in particle cooling processes. Non-thermally dominated sources, characterized by relatively low Lorentz factors in the accelerated electron population, produce MIR photons that are upscattered by inverse Compton processes, ultimately reaching the X-ray range \citep{Izumi2017ApJ...845L...5I}.

We implement this relation into the new X-ray module for CIGALE through the parameter $\alpha_\mathrm{IRX}$, defined as:

\begin{equation}
    \alpha_\mathrm{IRX} = \log \frac{L^\mathrm{int}_\mathrm{2-10~keV}}{L^\mathrm{nuc}_\mathrm{12~\mu m}}
\end{equation}

\begin{figure}[h!]
    \centering
    \includegraphics[width=1.0\linewidth]{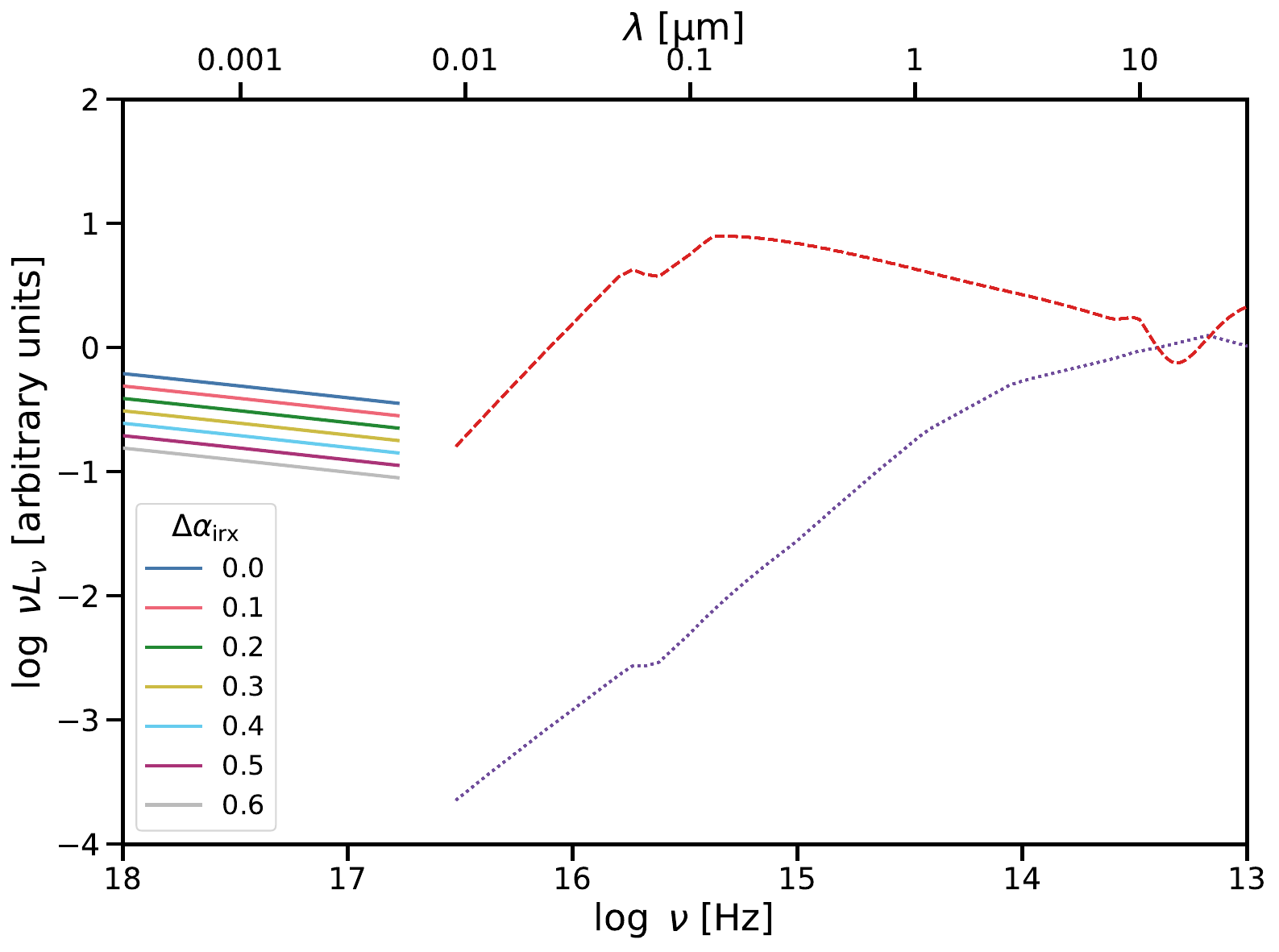}
    \caption{SED of two AGN for different $\alpha_\mathrm{IRX}$ normalized at 12 microns. The blue dotted and red dashed lines represent the UV-to-IR SED using $\delta_\mathrm{AGN}$ = 0 and 1, respectively. The solid lines are the X-ray SED, with different colors indicating different $\alpha_\mathrm{IRX}$ and a fixed power law index. Similarly to \citet{Yang20}, the ‘breaks’ between the two models are caused by the wavelength limit of the X-ray module, however, X-ray/UV observations are limited in these wavelengths.}
    \label{fig:alphairx-show}
\end{figure}

We have set $\alpha_\mathrm{IRX}$ with a default value of 0.3, and a grid going from 0.0 to 0.6. This choice is supported by the findings of \citet{Asmus15}, where their dispersion is measured at 0.33. The calculation of $L^\mathrm{nuc}\mathrm{12~\mu m}$ in CIGALE considers the sum of the three AGN components (accretion disk, polar dust, and toroid), while $L^\mathrm{int}\mathrm{2-10~keV}$ is derived from the intrinsic power-law fit for the AGN component. We assume the relationship holds at low luminosities because even at low accretion rates, both processes will be similarly affected due to the scarcity of UV photons. As with the previous X-ray CIGALE module, potential host contamination from X-ray binaries and host galaxy dust emissions will be accounted for in the grid models for SED fitting. Similarly, this new model also requires users to input absorption-corrected X-ray flux to derive intrinsic X-ray luminosity. \Cref{fig:alphairx-show} displays example SEDs of different $\alpha_\mathrm{IRX}$.

\begin{figure*}[t!]
    \centering
    \includegraphics[width=0.45\linewidth]{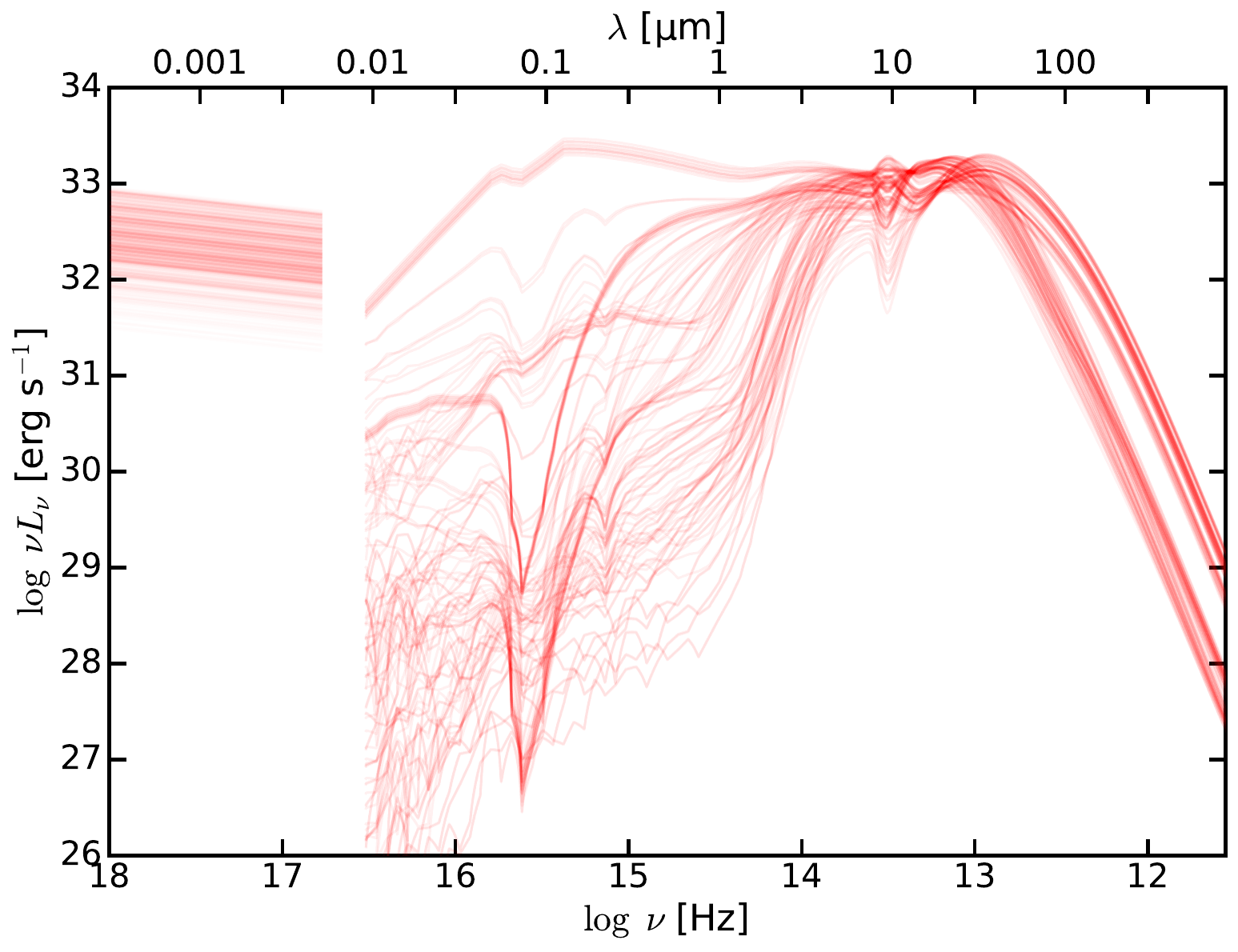}
    \includegraphics[width=0.45\linewidth]{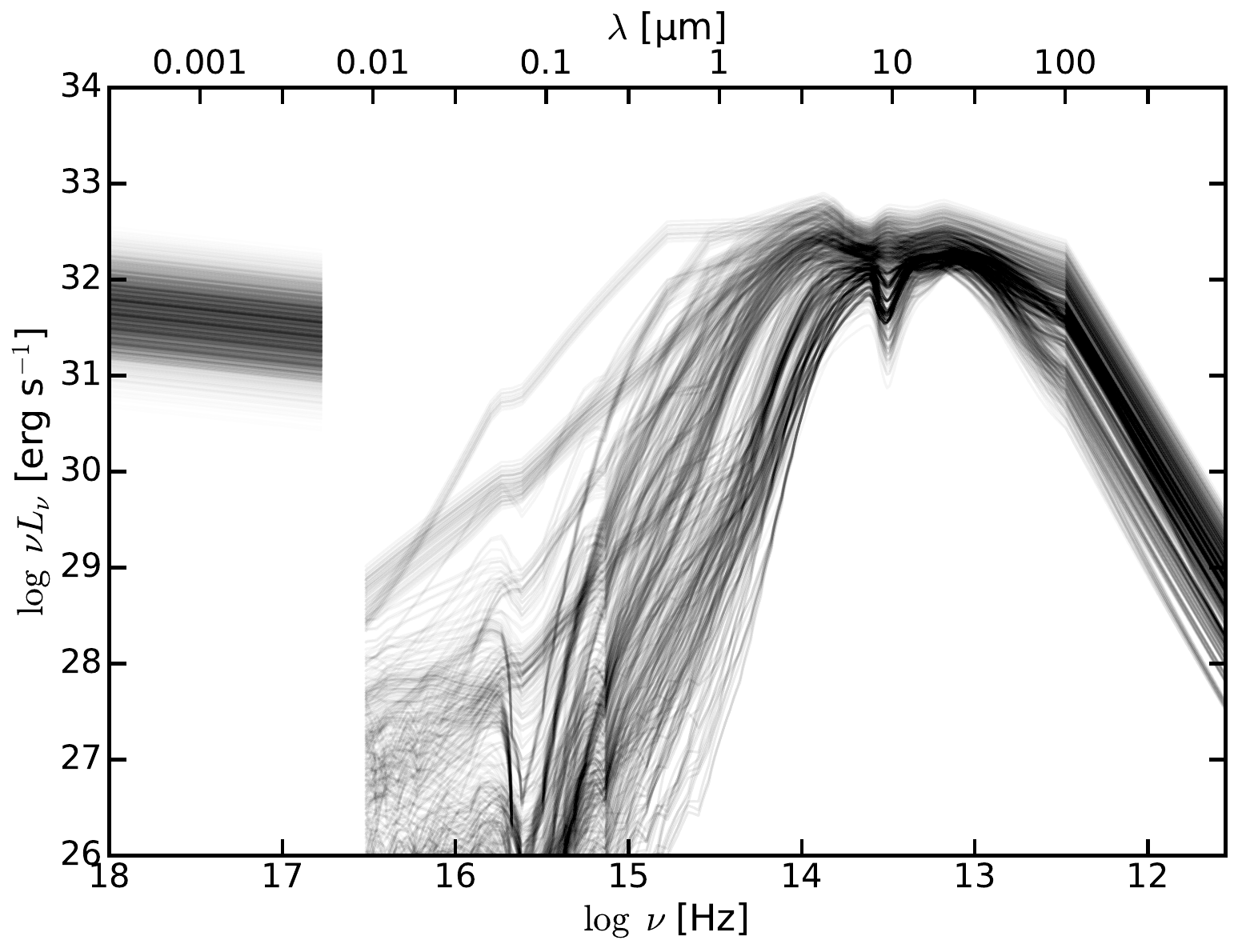}
    \caption{Gallery of AGN emission templates. On the left panel, the red lines are the observed AGN emission templates built with a fixed $\delta_\mathrm{AGN} = 1$ and our new $\alpha_\mathrm{IRX}$ variation. On the right panel, the black lines show a gallery of AGN templates built with varying $\delta_\mathrm{AGN}$ to lower values. These lower values of the new $\delta_\mathrm{AGN}$ produce a larger variation in UV slopes and IR behavior compared to those generated by the previous standard accretion disk model.}
    \label{fig:mock_seds}
\end{figure*}

These two refinements aim to provide a more precise and customized modeling strategy for LLAGN within the CIGALE framework. The code is accessible through the official CIGALE repository\footnote{This code will be included in a future release of Cigale. Until then, the current version is available on this branch: \url{https://gitlab.lam.fr/cigale/cigale/-/tree/alpha_irx?ref_type=heads}}. Incorporating a new X-ray module requires renaming the previous X-ray module \citep{Yang20}. To differentiate between the two, the previous \verb|xray| module will now be referred to as \verb|yang20|, while this new module will be called \verb|lopez24|.

The new central engine model can be selected within the AGN module and operates independently of the \verb|lopez24| X-ray module. However, we advise against using the new central engine model with low 
$\delta_\mathrm{AGN}$ values in conjunction with the \verb|yang20| module, as the $\alpha_\mathrm{ox}$--$L_\mathrm{2500\AA}$ relation in \verb|yang20| is not calibrated for these types of accretion, as discussed earlier.

Although initially developed for CIGALE, the underlying concept is flexible and can be adapted for use with other SED fitting codes. In the following subsection, we address the strengths, limitations and uncertainties of our module so future works can improve parameter accuracy and provide a more accurate representation of the physical processes occurring in LLAGN.

\subsection{Mock analysis}
\label{subsec:2.3}

We performed a mock analysis that demonstrates not only the expanded variety of AGN templates that CIGALE can now reproduce, but also the limitations of our module. A mock dataset of 365,904 galaxies was generated, with substantial variations in the AGN module, particularly focusing on the new parameters $\alpha_\mathrm{IRX}$ and $\delta_\mathrm{AGN}$. We employed the new progressive accretion model, allowing $\delta_\mathrm{AGN}$ to vary between 0 and 1, and $\alpha_\mathrm{IRX}$ to range from 0.0 to 0.6. Additionally, the AGN contribution to the total dust luminosity of the host galaxy (\fracAGN) was varied between 0.01 and 0.99. A wide range of AGN parameters were used, including variations in average edge-on optical depth, dust density gradient, torus cone angle, polar dust, and inclination. These parameters provided a comprehensive representation of potential AGN and LLAGN properties.

\begin{figure}[t!]
    \centering
    \includegraphics[width=0.8\linewidth]{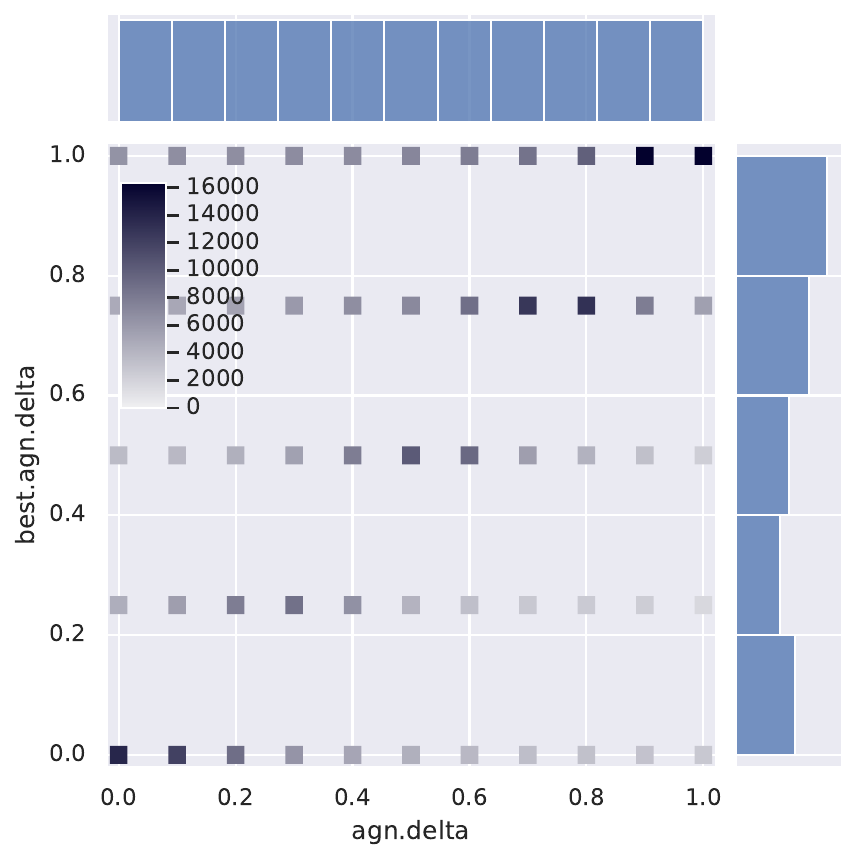}\\
    \includegraphics[width=0.8\linewidth]{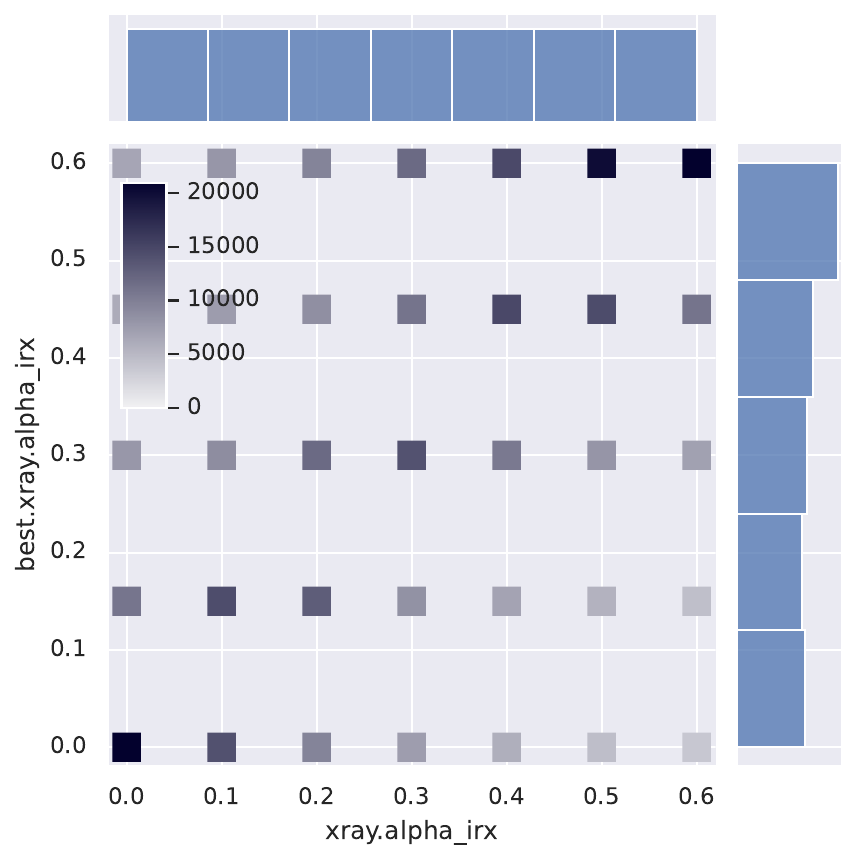}
    \caption{The top panel shows the true $\delta_\mathrm{AGN}$ parameter and its best-fit estimation after applying Gaussian errors and performing SED fitting. Most solutions fall along the 1:1 relation, indicating accurate parameter recovery. The x-axis histogram illustrates the uniform distribution of the true grid values, while the y-axis histogram is the distribution of the recovered properties. The bottom panel presents a similar analysis for $\alpha_\mathrm{IRX}$, showing the true values and their corresponding best-fit estimations.}
    \label{fig:mock_parameters}
\end{figure}

\begin{figure}[]
    \centering
    \includegraphics[width=1.\linewidth]{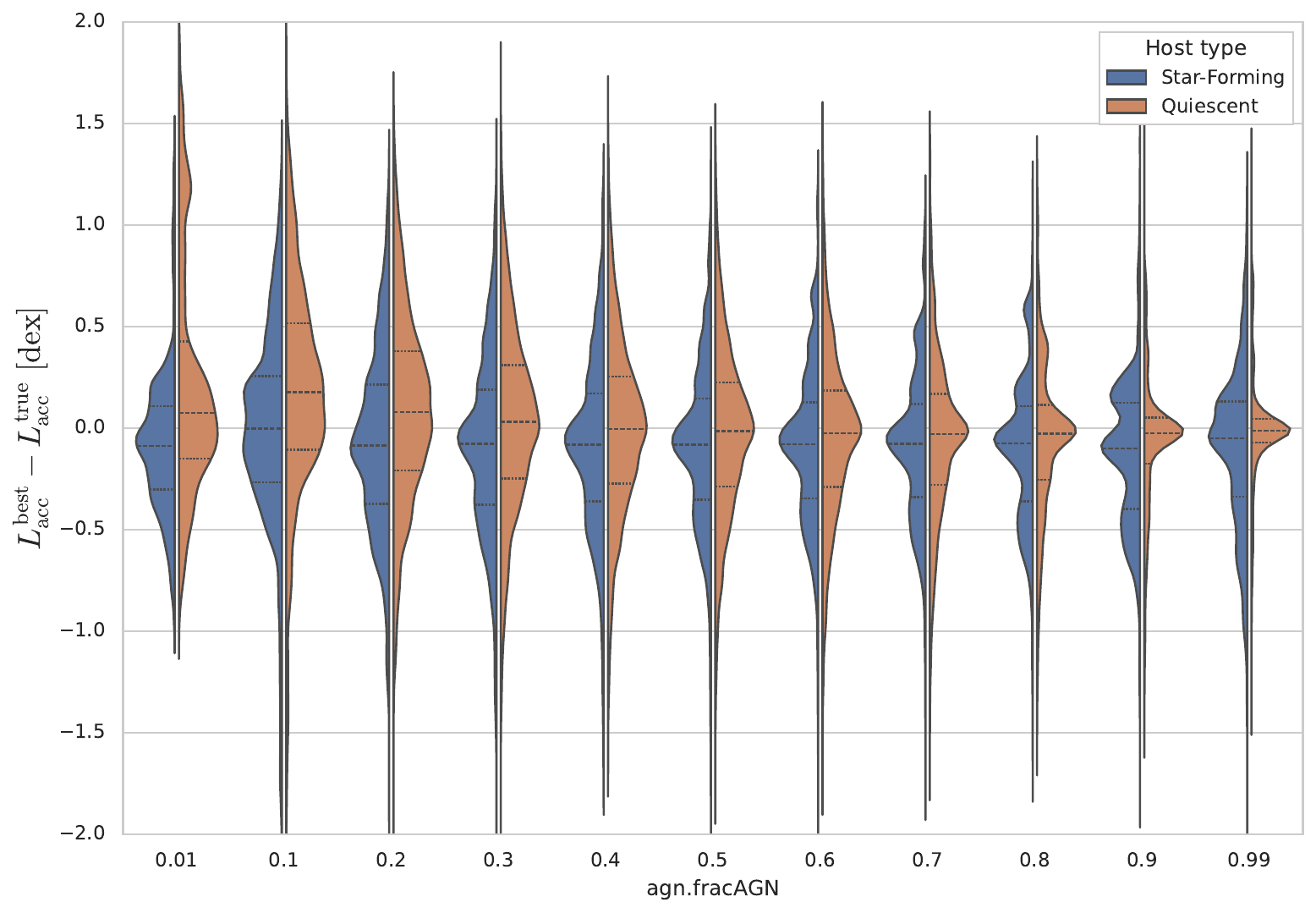}
    \caption{Violin plots showing the difference between the true AGN accretion power and the fitted value. Blue represents a star-forming host galaxy with a young stellar population, while orange represents a quiescent host galaxy with an old stellar population. Dashed lines inside the violins indicate the quartiles of each distribution.}
    \label{fig:violin}
\end{figure}

A representation of the variety in AGN emission produced by the mock analysis is shown in \Cref{fig:mock_seds}. In this figure, we compare the SEDs generated by the new module's central engine with those produced using only a Schartmann central engine. When assuming a standard accretion disk as the central engine, any modification in the UV-optical slope is attributed to absorption by the torus or polar dust, leading to re-emission of these components in the IR. Our model captures a broader range of UV slopes and IR behaviors, consistent with the underlying hypotheses of the new module. This comparison highlights the significant improvements in the module’s ability to generate a wide variety of AGN templates, including reduced UV emission, a more pronounced red bump at 1–10 microns, and diminished dust re-emission at 10–100 microns due to the UV field.

After generating the templates for AGN and host galaxy emission, we calculated the fluxes for the 2–10 keV and 0.5–2 keV X-ray bands, as well as fluxes from the GALEX NUV and FUV filters, SDSS ugriz bands, 2MASS JHKs bands, WISE bands 1–4, Herschel PACS at 70, 100, and 160 microns, and Herschel SPIRE at 250 microns. These mock flux densities were perturbed by adding noise, randomly sampled from a Gaussian distribution with a standard deviation based on the mean errors from \ref{subsec:3.1}. Once the mock fluxes were obtained, we performed SED fitting to determine if the true parameters could be accurately recovered.

A smaller grid of parameters was used during the fitting process to evaluate potential bias when the true grid did not perfectly match the fitted one. We assumed the host galaxies had known characteristics. \Cref{fig:mock_parameters} shows the recovered parameters $\delta_\mathrm{AGN}$ and $\alpha_\mathrm{IRX}$ for a quiescent galaxy (i.e., an old stellar population, low star formation rates, and varying attenuation values). Tests performed on galaxies with higher star formation rates revealed no significant differences. The results indicate that most of the recovered parameters fall along the 1:1 relation, demonstrating the reliability of the new module in accurately recovering these values. The median absolute deviation between the true and estimated parameters was 0.2 for $\delta_\mathrm{AGN}$ and 0.15 for $\alpha_\mathrm{IRX}$, with mean squared errors of 0.16 and 0.06, respectively. The histogram of the true values shows a uniform distribution of grid points, while the histogram of the recovered values shows a similar distribution, with a slight preference for higher values in both parameters.

Finally, and most importantly, we tested the accuracy of the \accretion, which represents the AGN 'disk' luminosity averaged over all directions. This parameter is typically used to derive Eddington ratios, as it is more representative of accretion than other luminosity measures that include torus re-emission. \Cref{fig:violin} illustrates the distribution of the difference between the best-fitted value and the true value from our mock analysis, across a range of \fracAGN. As expected, when the AGN dominates, the accretion luminosity is better constrained. For quiescent galaxies, as \fracAGN\ decreases, the accretion luminosity exhibits more variance, but the median value consistently remains close to zero. At extremely low fractions, such as 0.01, while most galaxies show a difference close to zero, a small number of models display deviations of up to 1 dex. For star-forming galaxies, the distribution remains relatively uniform across the full range of \fracAGN, without significant differences at lower or higher \fracAGN. Overall, for both types of host galaxies and across most \fracAGN\ values, more than 50\% of the models have a difference within 0.5 dex.

\section{Data and Fitting}
\label{IRX-sec:3}

This section describes the two samples used to evaluate the new CIGALE module presented in this work. The primary sample discussed in \cref{subsec:3.1} comprises 50 local galaxies classified as Seyferts and LINERs. This sample will assess the LLAGN properties and validate the module's performance in the low luminosity regime. As a secondary sample (\cref{subsec:3.2}), we use the COSMOS and SDSS QSO samples employed in previous CIGALE studies. These sources will be used to confirm the validity of our $L_\mathrm{X}$--$L_\mathrm{12\mu m}$ implementation in the mid-luminosity regime. Finally, \cref{subsec:3.3} discusses the fitting methodology and goodness of the fit.

\subsection{LLAGN Sample}
\label{subsec:3.1}

\begin{figure}[t!]
    \centering
    \includegraphics[width=1\linewidth]{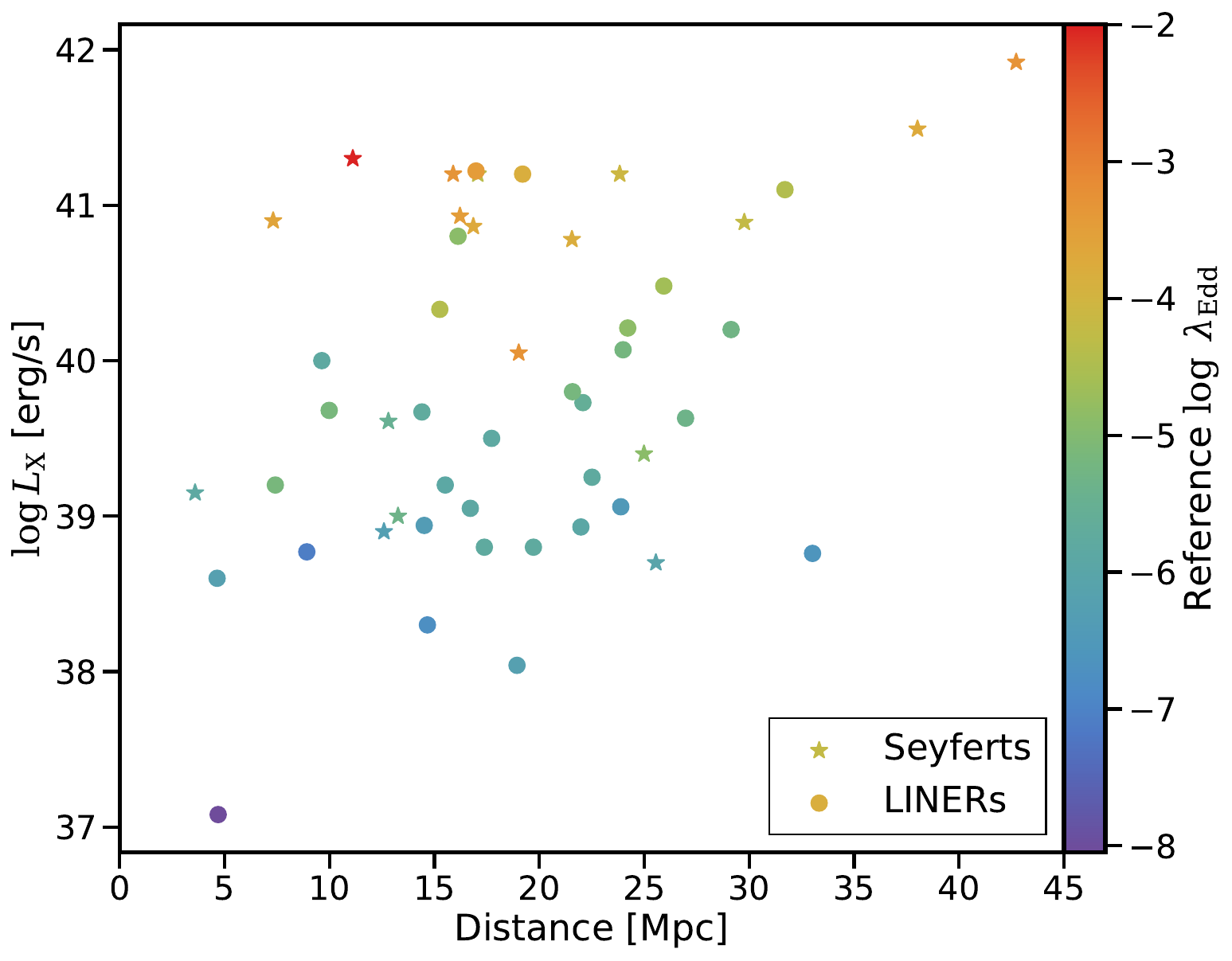}
    \caption[Properties of the LLAGN sample]{X-ray luminosity vs. distance for the sample of 50 LLAGN, with Eddington ratios color-coded. Most galaxies are located within 40 Mpc, and all have Eddington ratios $\log \lambda_\mathrm{Edd}$\,$<$\,$-2$, indicating low accretion rates consistent with RIAFs.}
    \label{fig:LLAGN-sample-edd}
\end{figure}

We compiled a sample of 50 local galaxies, each classified as either LINER or Seyfert based on emission line diagnostics \citep{Ho1997ApJS..112..315H}, all of which feature central X-ray detections with $L_\mathrm{X}<10^{42}\,\mathrm{erg\,s^{-1}}$. These galaxies are characterized by low accretion rates, as indicated by their Eddington ratios\footnote{The reference bolometric luminosities (used to calculate Eddington ratios for sample selection and future comparison in \cref{fig:validation} with our results) were primarily derived using bolometric corrections from X-ray luminosities, adopting $k_\mathrm{X}=15.8$ as per \citet{Ho2009ApJ...699..626H}, or through SED fitting, as described by \citet{Eracleous2010ApJS..187..135E} and \citet{Nemmen14}.}. \Cref{fig:LLAGN-sample-edd} shows the X-ray luminosity, Eddington ratio, and distance of our sample, where most galaxies are located within 40 Mpc. All the LLAGN have $\log \lambda_\mathrm{Edd}$\,$<$\,$-2$, which makes them candidates for radiatively inefficient accretion flows (RIAFs), suggesting that the central engines may operate under this accretion mode \citep{Ho2009ApJ...699..626H}.

\begin{figure}[h!]
    \centering
    \includegraphics[width=1\linewidth]{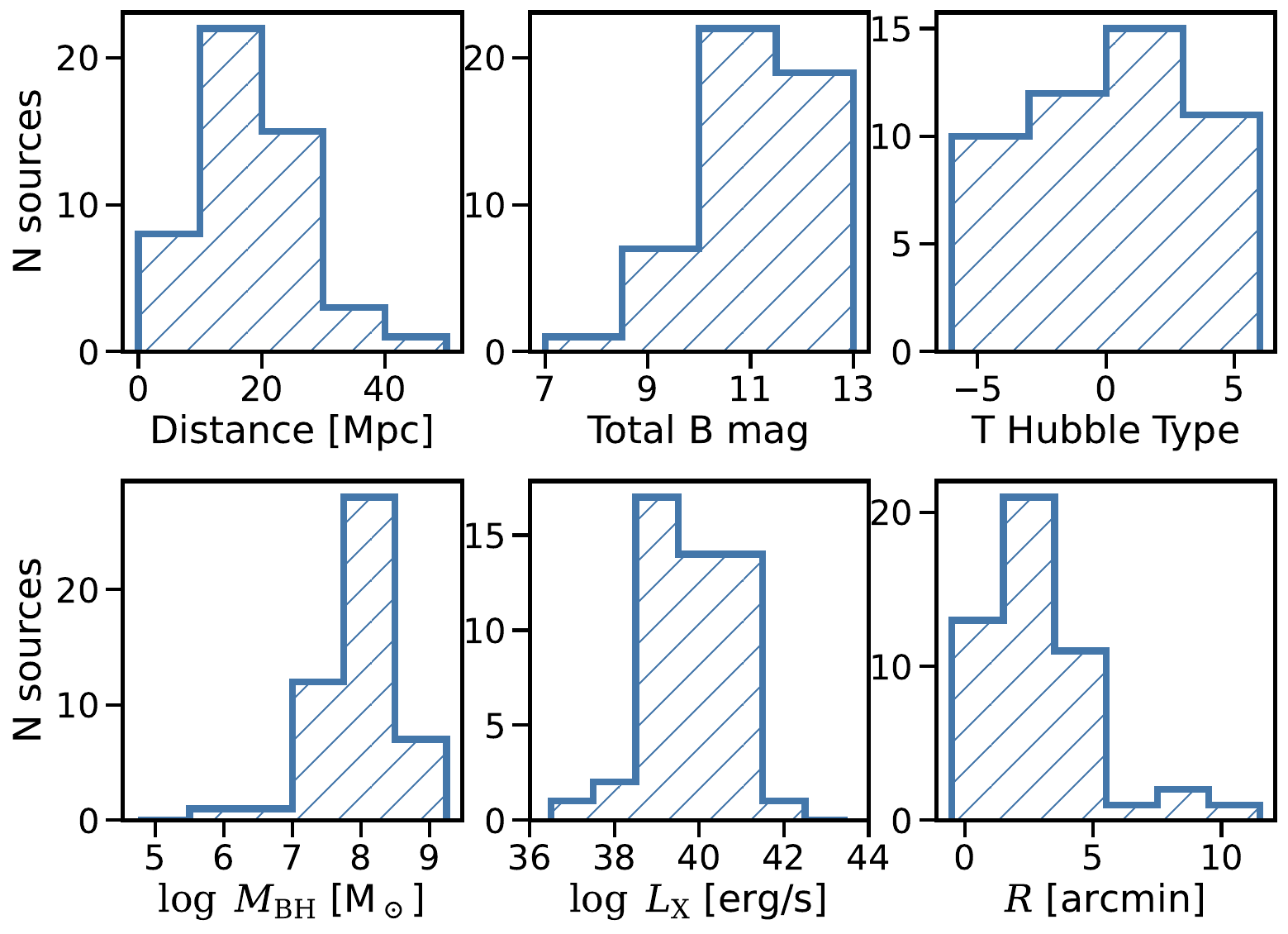}
    \caption[Properties of the LLAGN sample]{Distributions illustrating the fundamental properties of the host galaxies and their associated supermassive black holes in the LLAGN sample considered in this work.}
    \label{fig:LLAGN-sample}
\end{figure}

Most of these galaxies have an angular size of less than five arcminutes in radius and span a wide range of morphologies, with Hubble types ranging from -5 to 5. The SMBH masses, retrieved from the literature, mostly fall within the range of $7$\,$<$\,$\log~M_\mathrm{BH}$\,$<$\,$9$, though the sample includes several galaxies hosting less massive SMBHs. A comprehensive list of the galaxies and their properties is provided in \Cref{tab:llagn-sample}, with the distributions of these parameters shown in \Cref{fig:LLAGN-sample}.

\begin{table}
    \centering
    \smaller
    \begin{tabular}{ccccccc}
    \hline\hline
     ID &   RA &   Dec &  z &  Dist &  $\log L_\mathrm{X}$ &  Ref \\
      &    &    &   & [Mpc] &  [erg/s] &   \\
    \hline
     NGC1052 &   40.269 &  -8.2558 &  0.00496 &  19.19 &    41.20 &       2 \\
     NGC2273 &  102.536 &  60.8457 &  0.00613 &  29.76 &    40.89 &       1 \\
     NGC2685 &  133.894 &  58.7343 &  0.00295 &  13.26 &    39.00 &       2 \\
     NGC2655 &  133.909 &  78.2234 &  0.00467 &  23.82 &    41.20 &       2 \\
     NGC2768 &  137.906 &  60.0373 &  0.00451 &  22.07 &    39.73 &       1 \\
     NGC2787 &  139.826 &  69.2033 &  0.00232 &   7.41 &    39.20 &       1 \\
     NGC2841 &  140.511 &  50.9766 &  0.00211 &  14.51 &    38.94 &       1 \\
     NGC3031 &  148.888 &  69.0652 & -0.00013 &   3.59 &    39.15 &       1 \\
     NGC3079 &  150.491 &  55.6799 &  0.00368 &  19.01 &    40.05 &       1 \\
     NGC3147 &  154.223 &  73.4007 &  0.00934 &  42.72 &    41.92 &       1 \\
     NGC3185 &  154.410 &  21.6883 &  0.00410 &  24.98 &    39.40 &       2 \\
     NGC3190 &  154.526 &  21.8326 &  0.00437 &  23.98 &    40.07 &       1 \\
     NGC3193 &  154.603 &  21.8940 &  0.00460 &  33.01 &    38.76 &       1 \\
     NGC3414 &  162.817 &  27.9750 &  0.00490 &  25.92 &    40.48 &       1 \\
     NGC3516 &  166.697 &  72.5685 &  0.00883 &  38.01 &    42.49 &       1 \\
     NGC3718 &  173.145 &  53.0679 &  0.00331 &  16.98 &    41.22 &       1 \\
     NGC3898 &  177.313 &  56.0843 &  0.00385 &  21.97 &    38.93 &       1 \\
     NGC3945 &  178.306 &  60.6755 &  0.00427 &  21.57 &    39.80 &       1 \\  
     NGC3953 &  178.454 &  52.3266 &  0.00350 &  18.93 &    38.04 &       1 \\
     NGC3982 &  179.117 &  55.1252 &  0.00371 &  21.55 &    40.78 &       5 \\
     NGC4013 &  179.630 &  43.9468 &  0.00277 &  19.71 &    38.80 &       1 \\
     ...     &  ... &  ... &  ... &  ... &    ... &       ... \\
    \hline
    \end{tabular}
    \caption[LLAGN Sample Properties]{Key properties of galaxies in the LLAGN sample. Distances are given in Mpc.  X-ray luminosities are rest-frame and absorption-corrected, expressed in erg s$^{-1}$. The table includes references for the X-ray luminosities: 1 for \citet{Williams22}, 2 for \citet{GonzalezMartin09}, 3 for \citet{OsorioClavijo23}, 4 for \citet{Yun22}, 5 for \citet{Kammoun20}, 6 for \citet{Cappi06}, and 7 for \citet{Masini22}.}
    \label{tab:llagn-sample}
\end{table}

\begin{figure*}[t!]
    \centering
    \includegraphics[width=1\linewidth]{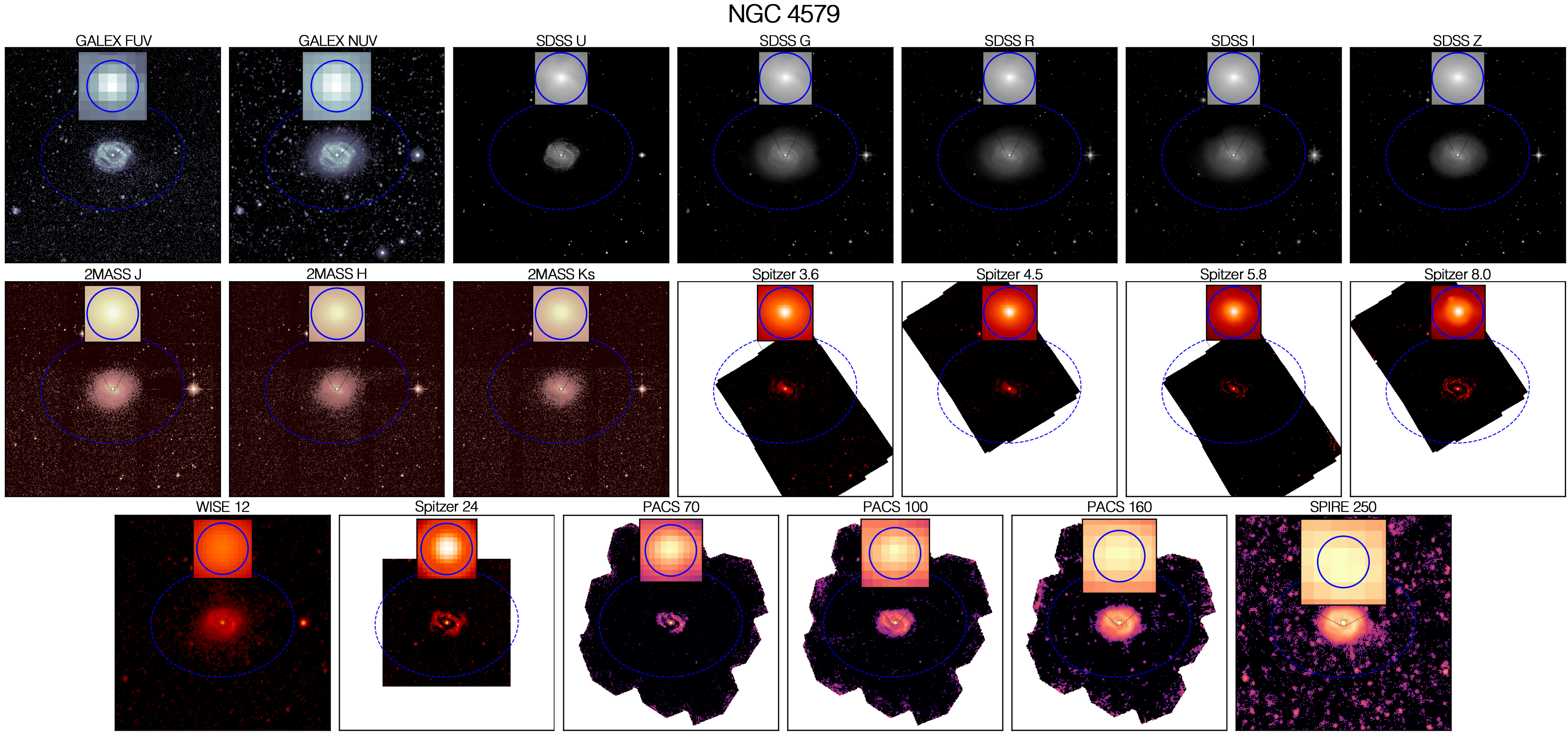}
    \caption[Example of our multiwavelength photometry]{
    Example of forced aperture photometry with a 9-arcsecond radius, executed across all wavelengths from GALEX NUV to Herschel SPIRE 250 microns. The blue dashed ellipse delineates the galaxy's boundary as identified by isophotes from DustPedia. An inset offers a detailed view of the 9-arcsecond aperture region utilized in this analysis, showcased as a blue solid circle.}
    \label{fig:photometry}
\end{figure*}

In particular, nine galaxies were classified as Seyfert galaxies with low Eddington ratios: NGC2273, NGC3185, NGC3516, NGC3982, NGC4051, 
NGC4258, NGC4565, NGC4725, NGC5273. While the accretion regime is already inefficient, a standard accretion disk is not probable; sixty percent of them have independent evidence of showing radiatively inefficient accretion. Some galaxies show hints of having an ADAFs scenarios like NGC3516 \citep{Cao2014MNRAS.444L..20C}, NGC4258 \citep{Doi2005MNRAS.363..692D,Szanecki2023MNRAS.521.2215S}, NGC4565 \citep{Doi2005MNRAS.363..692D,LiuWu2013ApJ...764...17L} or NGC 5273 \citep{Giroletti2009ApJ...706L.260G}. NGC4051 are discussed to be a pure accretion flow scenario \citep{Peterson2000ApJ...542..161P,Meyer2011A&A...527A.127M}, or a truncated accretion disk with a comptonizated region \citep{Zhang2006ChJAA...6..165Z}. 

The rest of the galaxies are classified as LINERs, from which all show evidence of hosting an AGN, mainly through the ionization power of its central X-ray emission \citep[e.g., NGC2655,][]{Reynaldi2020MNRAS.499.5107R} or through radio nuclear emission with compact jets \citep[e.g.,][]{Baldi18LemmingsI,Baldi21LemmingsII}. For sixty percent of them, there are independent hints of an ADAF scenario or non standard accretion disk scenario: NGC1052 \citep{Yuan2009ApJ...703.1034Y,Falocco2020A&A...638A..67F}, NGC2787 \citep{Doi2005MNRAS.363..692D,Pellegrini2005ApJ...624..155P, LiuWu2013ApJ...764...17L}, NGC2841 \citep{Pellegrini2005ApJ...624..155P}, NGC3031 \citep{Doi2005MNRAS.363..692D,Nemmen14}, NGC3079 \citep{Chang2002AJ....124.1948C}, NGC3147 \citep{Doi2005MNRAS.363..692D, LiuWu2013ApJ...764...17L}, NGC3414 \citep{Wojtowicz2023ApJ...944..195W}, NGC3718 \citep{Doi2005MNRAS.363..692D}, NGC4111 \citep{Hauschild2022MNRAS.512.2556H}, NGC4125 \citep{Pellegrini2005ApJ...624..155P}, NGC4138 \citep{LiuWu2013ApJ...764...17L}, NGC4203 \citep{Doi2005MNRAS.363..692D,Yuan2009ApJ...703.1034Y,LiuWu2013ApJ...764...17L}, NGC4261 \citep{Yuan2009ApJ...703.1034Y,Nemmen14,Pellegrini2005ApJ...624..155P}, NGC4374 \citep{Doi2005MNRAS.363..692D, Pellegrini2005ApJ...624..155P}, 
NGC4457 \citep{Nemmen14}, NGC4486  \citep{Doi2005MNRAS.363..692D,Nemmen14, Pellegrini2005ApJ...624..155P}, NGC4494  \citep{Nemmen14}, NGC4552 \citep{Doi2005MNRAS.363..692D,Nemmen14, Pellegrini2005ApJ...624..155P}, NGC4579  \citep{Doi2005MNRAS.363..692D,Nemmen14,Yuan2009ApJ...703.1034Y}, NGC4594 \citep{Nemmen14,Pellegrini2005ApJ...624..155P,LiuWu2013ApJ...764...17L}, NGC4698 \citep{Tran2011ApJ...726L..21T}, and NGC4736 \citep{Nemmen14,LiuWu2013ApJ...764...17L}.

The primary X-ray catalogs used for compiling the sample include \citet{Williams22}, \citet{OsorioClavijo23}, and \citet{GonzalezMartin09}. These catalogs provide spectral fits to data from the Chandra and XMM-Newton telescopes, utilizing a variety of models. These range from the classic power-law model, attenuated by host galaxy and Milky Way obscuration, to more complex models incorporating features like the Iron K$\alpha$ fluorescence line, absorber models, reflection components, and soft X-ray emitters such as ionized plasma. The spectral fits were performed for detections meeting a 3-$\sigma$ threshold. Our focus is on the intrinsic emission from the corona or ADAF, which is characterized by power-law spectra. The extracted power-law index ($\Gamma$) and normalization values were used to compute intrinsic X-ray fluxes in both the soft (0.5–2 keV) and hard (2–10 keV) X-ray bands.

For specific sources, we utilized data from works that incorporate more detailed reflection components, complex thermal elements, or torus models.  These work, usually characterize the intrinsic power-law emission using data from telescopes operating at higher energies, such as Swift-BAT and NuSTAR, which are less affected by obscuration. For detailed information on individual sources, we refer to \citet{Yun22} for NGC5033, \citet{Kammoun20} for NGC3982, and \citet{Masini22} for NGC4258. Additionally, for NGC4051, 
and NGC4725, the spectral fittings are taken from \citet{Cappi06}. 

For each galaxy, we conducted photometry spanning UV to FIR bands using calibrated images from the DustPedia database\footnote{\url{http://dustpedia.astro.noa.gr/Data}}. We performed forced photometry in a 9-arcsecond radius aperture centered on each galaxy (refer to \cref{fig:photometry} for an example). This aperture covered GALEX NUV and FUV filters, ugriz SDSS bands, JHKs bands from 2MASS, Spitzer IRAC1-4 and WISE 1-4, Spitzer MIPS 24-70 microns, 100, Herschel PACS 70-100-160 microns, and Herschel SPIRE 250 microns. The 9-arcsecond radius was deliberately chosen to align with the diameter required to encompass the Full Width at Half Maximum (FWHM) of the Point Spread Function (PSF) for each filter in our analysis. Upper limits at $3\sigma$ were set for non-detected sources.

Using the Photutils library version 1.9\footnote{\url{https://photutils.readthedocs.io/}}, we utilized a two-dimensional modeling approach to perform background calculations. To determine the background, we also implemented a sigma clipping technique and, where feasible, masked the galaxy using a mask derived from the largest isophote in the DustPedia database. Additionally, we applied aperture corrections and corrected for Milky Way extinction, utilizing the dust extinction function from \citet{Fitzpatrick1999PASP..111...63F} and a sky dust map provided by \citet{SFDMaps}.

\subsection{AGN Samples}
\label{subsec:3.2}

To verify the robustness of the $L_\mathrm{X}$--$L_\mathrm{12\mu m}$ implementation described in \cref{subsec:2.2.2}, we utilized mid-luminosities AGN samples from well-established datasets. Although the extrapolation of the $L_\mathrm{X}$--$L_\mathrm{12\mu m}$ to high-luminosity sources varies by an order of magnitude \citep{Stern2015ApJ...807..129S}, the relation remains reliable for $L_\mathrm{X}<10^{45}\ \mathrm{erg\ s^{-1}}$ \citep{Asmus15}. Specifically, we focused on the COSMOS and SDSS QSO samples referenced in previous CIGALE publications \citep{Yang20,Yang22}, which enable direct comparisons with the established X-ray module. These studies determined the bolometric luminosity of AGN using the $\alpha_\mathrm{OX}$ prior, which we adopted as a benchmark to assess the robustness of our new $\alpha_\mathrm{IRX}$ prior.

For both the SDSS and COSMOS samples, the SED fitting methodology used the same photometry and parameters established in \citet{Yang20}, retaining the classic accretion disk model from \citet{Schartmann05} ($\delta_\mathrm{AGN}=1$) and incorporating only the new X-ray module with the $\alpha_\mathrm{IRX}$ for comparative analysis (see \cref{subsec:4.1}).

We provide a brief summary of the AGN samples; for detailed information, please refer to \citet{Yang20}. The SDSS sample consists of Type I AGN that are optically selected from the DR14 quasar catalog \citep{Paris2018A&A...613A..51P}, including 1986 AGNs detected in the 2–12 keV X-ray band with XMM-Newton at a significance level greater than 3$\sigma$. Galactic extinction corrections were applied, and X-ray data were obtained from archival observations in the 3XMM catalog \citep{Rosen2016A&A...590A...1R}. The COSMOS sample, derived from the COSMOS-Legacy survey \citep{Civano2016ApJ...819...62C}, comprises 590 X-ray-selected AGNs detected in the 2–10 keV band at significance levels greater than 3$\sigma$. Photometric data were sourced from the COSMOS2015 catalog \citep{Laigle2016ApJS..224...24L}, covering 14 broad optical and infrared bands, and where available, spectroscopic information was compiled from \citet{Marchesi2016ApJ...817...34M}.

\subsection{LLAGN Fitting}
\label{subsec:3.3}

The SED fitting process for the LLAGN sample was conducted using a combination of CIGALE modules, comprehensively modeling both AGN and host galaxy emissions. The host galaxy characterization was performed using a \citet{2003PASP..115..763C} initial stellar mass function (IMF) with two possible metallicity values, a delayed star formation history for the stellar population and accounting for dust extinction. Dust emissions were effectively modeled using the Themis module \citep{Jones2017A&A...602A..46J}. 

For the AGN component, we modeled seed photons using our novel progressive accretion approach, as detailed in \cref{subsec:2.2.1}, alongside the new link between X-ray and UV-to-IR AGN emission described in \cref{subsec:2.2.2}. It is essential to clarify that the \verb|yang20| module was unable to fit AGN solutions for any of the 50 LLAGN included in the sample. The \verb|yang20| module generates all possible combinations of models between the X-ray and UV-to-IR AGN emission and assigns a chi-square value of NaN to models where the $\alpha_\mathrm{ox}$ relation exceeds a defined maximum dispersion. While it might be possible to obtain AGN solutions by adjusting the maximum deviation from the $\alpha_\mathrm{ox}$ relation, our tests revealed that permitting all levels of dispersion (maximum delta from the $\alpha_\mathrm{ox}$ relation = 0) rendered these values ineffective for constraining solutions, thereby reducing their utility in SED fitting.

Given that our LLAGN sample exhibited strong evidence of being in a radiatively inefficient accretion mode, we set the $\delta_\mathrm{AGN}$ to lower values (0, 0.25, 0.5), allowing a wide range of AGN fractions, predominantly centered on lower values (0.001 to 0.3). The toroid's opening angle was set at $10^\circ$ and $70^\circ$, achieving a covering factor of 80\% to better model the type of obscuration typical in low-accreting sources \citep{Ricci2017ApJS..233...17R}, while also accommodating the possibility of an almost disappearing torus \citep{Elitzur2008NewAR..52..274E}. Viewing angles were adjusted to allow a direct view of the central engine ($i=10^\circ$) or complete cover ($i=80^\circ$), with intermediate angles tested, though they showed no significant impact on the results. Additionally, the X-ray module was employed to model emissions from the host galaxy, accounting for emissions from X-ray binaries and the AGN X-ray power law across a range of power indices. A detailed list of the parameters used in the CIGALE input is presented in Table \ref{table:06cigale-input}.

\begin{table*}[p!]
\caption{Parameters and values for the modules used with CIGALE for the LLAGN sample.} 
\centering
\setlength{\tabcolsep}{1.mm}
\begin{tabular}{cc}
\hline
Parameter &  Model/values\\
\hline\\[-1.5ex]
\multicolumn{2}{c}{Star formation history: delayed model} \\[0.5ex]
Age of the main population & 3000, 5000, 7000, 9000, 11000 Myr  \\
e-folding time & 100, 500, 1000, 3000 Myr \\ 
Burst stellar mass fraction & 0.0 \\
\hline\\[-1.5ex]
\multicolumn{2}{c}{Simple Stellar population: Bruzual \& Charlot (2003)} \\[0.5ex]
Initial Mass Function & Chabrier (2003)\\
Metallicity & 0.008 (LMC), 0.02 (Solar) \\
\hline\\[-1.5ex]
\multicolumn{2}{c}{Dust extinction} \\[0.5ex]
Dust attenuation recipe & modified Calzetti et al. (2000)\\
E(B-V)$_{lines}$ &  0.1, 0.3, 0.5, 0.7, 0.9\\
E(B-V)$_{factor}$ & 0.1, 0.44, 0.7\\
Power-law slope modifying the attenuation curve &  -0.4, 0.0\\
\hline\\[-1.5ex]
\multicolumn{2}{c}{Dust emission: Themis (Jones et al. 2017)} \\[0.5ex]
 $q_\mathrm{HAC}$ & 0.02, 0.17, 0.24\\
$U_\mathrm{min}$ & 0.1, 1.0, 10.0, 50.0\\
$\alpha$ slope in $dM_{dust}\propto U^{-\alpha}dU$ & 2.1, 2.5, 2.9 \\
$\gamma$ & 0.4\\
\hline\\[-1.5ex]
\multicolumn{2}{c}{AGN module: SKIRTOR} \\[0.5ex]
Torus optical depth at 9.7 microns $\tau _{9.7}$ & 3, 11 \\
Torus opening angle  &$10^{\circ}$, $70^{\circ}$\\
Viewing angle  &$10^{\circ}\,\,\rm{(Type\,\,I)},80^{\circ}\,\,\rm{(Type\,\,II)}$ \\
Disk spectrum  & ADAF + Truncated disk \\
Delta parameter for the truncated disk & 0, 0.25, 0.5\\
AGN fraction &  0.001, 0.01, 0.05, 0.1, 0.2, 0.3 \\
$E(B-V)$ of polar dust & 0.0\\
\hline\\[-1.5ex]
\multicolumn{2}{c}{X-ray module: this paper} \\[0.5ex]
AGN photon index $\Gamma$ & 1.6, 2.0, 2.5, 3.0, 3.5 \\
$\alpha_{IRX}$ & 0, 0.15, 0.3, 0.45, 0.6\\
Deviation from expected LMXB and HMXB & 0., 0. \\
\hline\\[-1.5ex]
Total number of models per redshift & 671,846,400 \\[0.5ex]
\hline
\label{table:06cigale-input}
\end{tabular}
\end{table*}

Galaxies with fits resulting in $\chi^2_\mathrm{red}>5$ were excluded from the analysis, leading to the exclusion of only three galaxies out of the initial 50: NGC4013, NGC4826, and NGC5866. NGC4013 and NGC5866 present challenges due to their dusty nature and pronounced side-view orientation, complicating AGN detection and subsequent SED fitting. Specifically, NGC4013 experiences additional complexity due to star contamination near its center. Although NGC4826 has no side-view orientation, it is characterized by a prominent absorbing dust lane.

\begin{figure}[h!]
    \centering
    \includegraphics[width=0.75\linewidth]{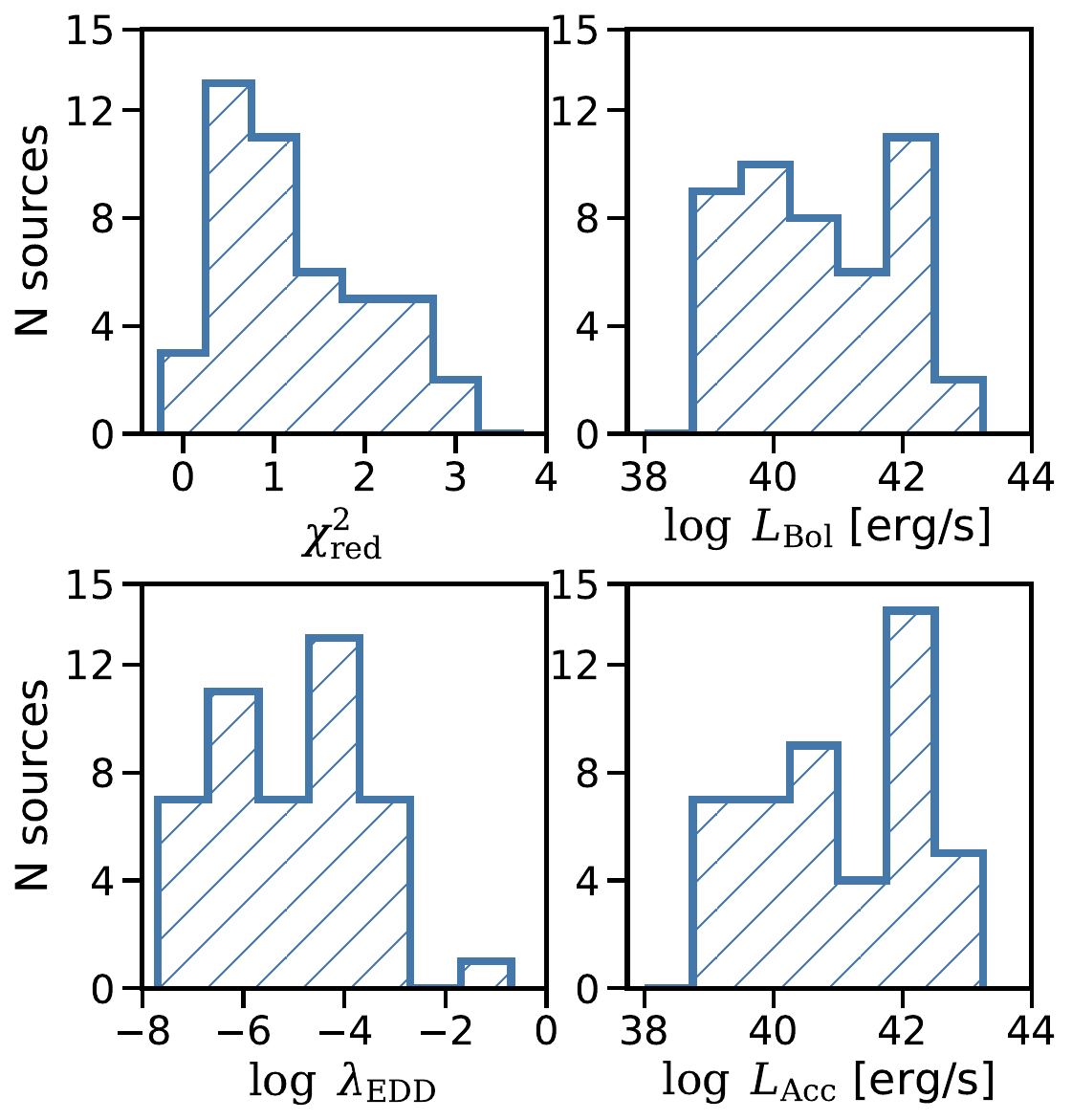}
    \caption[Distribution of physical parameters from the SED fitting]{Distribution of reduced chi-square ($\chi^2_\mathrm{red}$) values and key AGN properties derived from the SED fitting of the LLAGN sample. Reduced chi-square values near 1 signify a good fit quality. This figure also shows the values for AGN properties, such as bolometric luminosity ($L_\mathrm{Bol}$) and accretion luminosity ($L_\mathrm{Acc}$), from which we calculated the Eddington ratio ($\lambda_\mathrm{EDD}$) for each source.}
    \label{fig:LLAGN-results}
\end{figure}

\begin{figure*}[t!]
    \centering
    \includegraphics[width=0.49\linewidth]{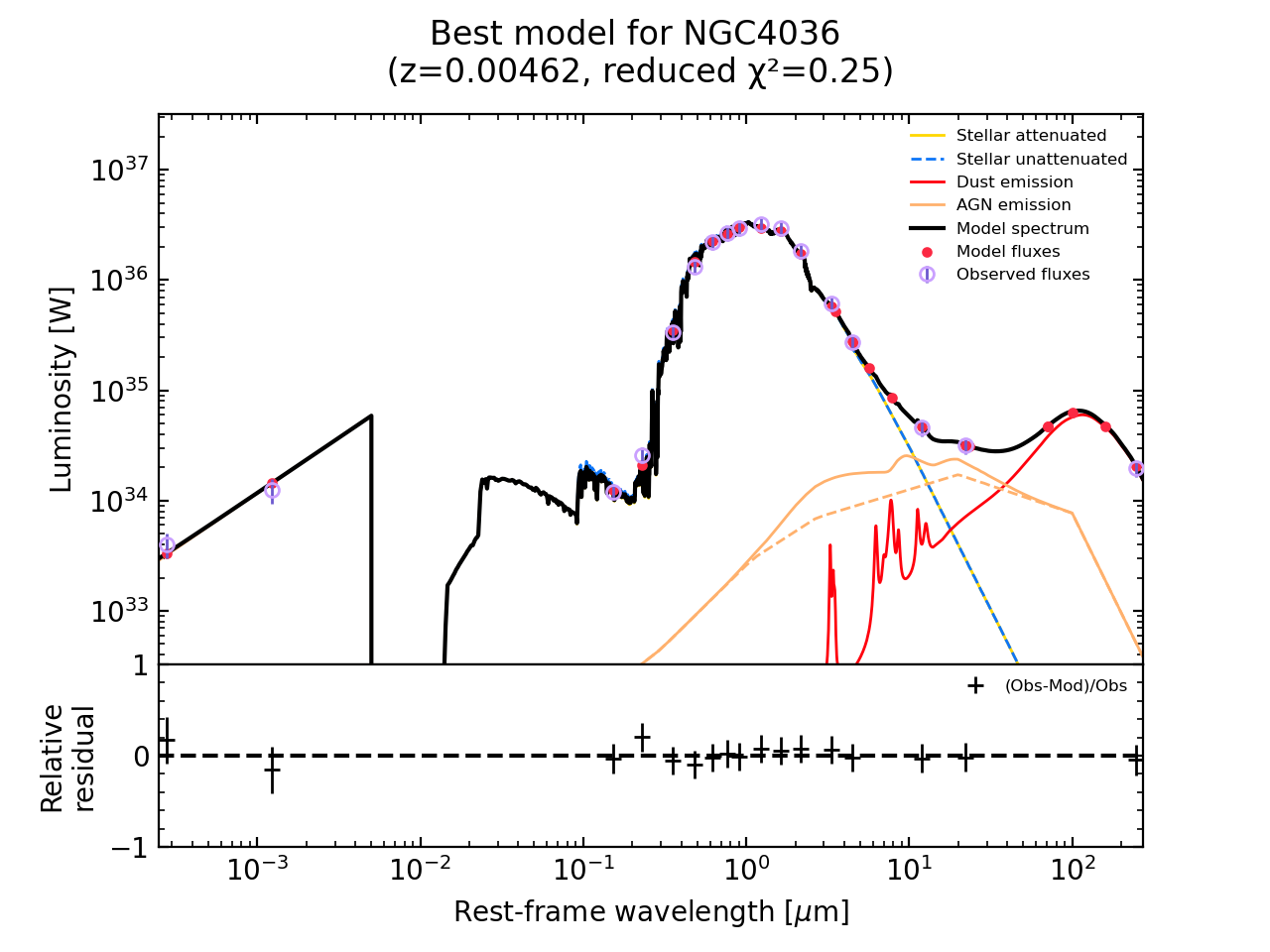}
    \includegraphics[width=0.49\linewidth]{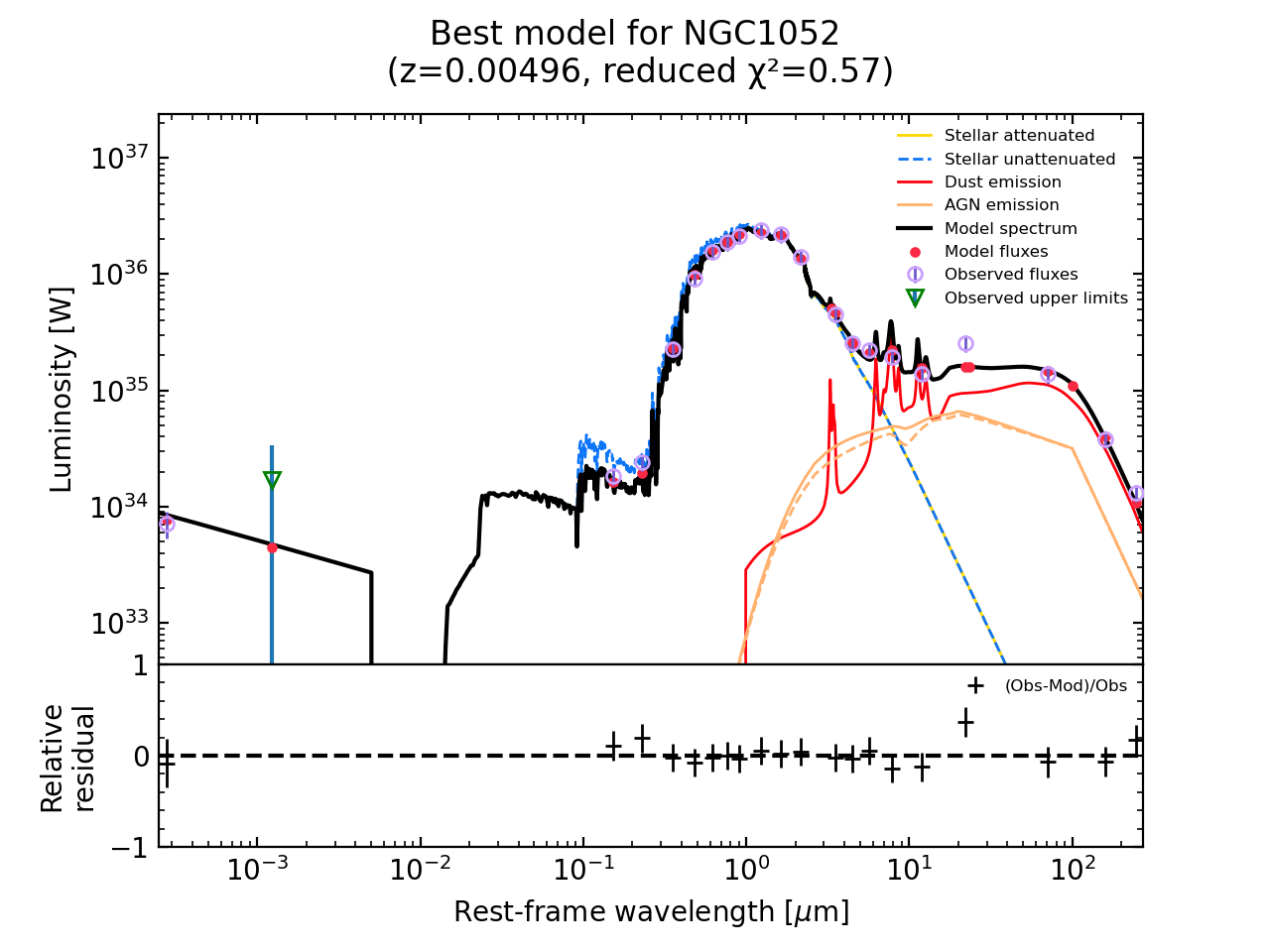}
    \\
    \includegraphics[width=0.49\linewidth]{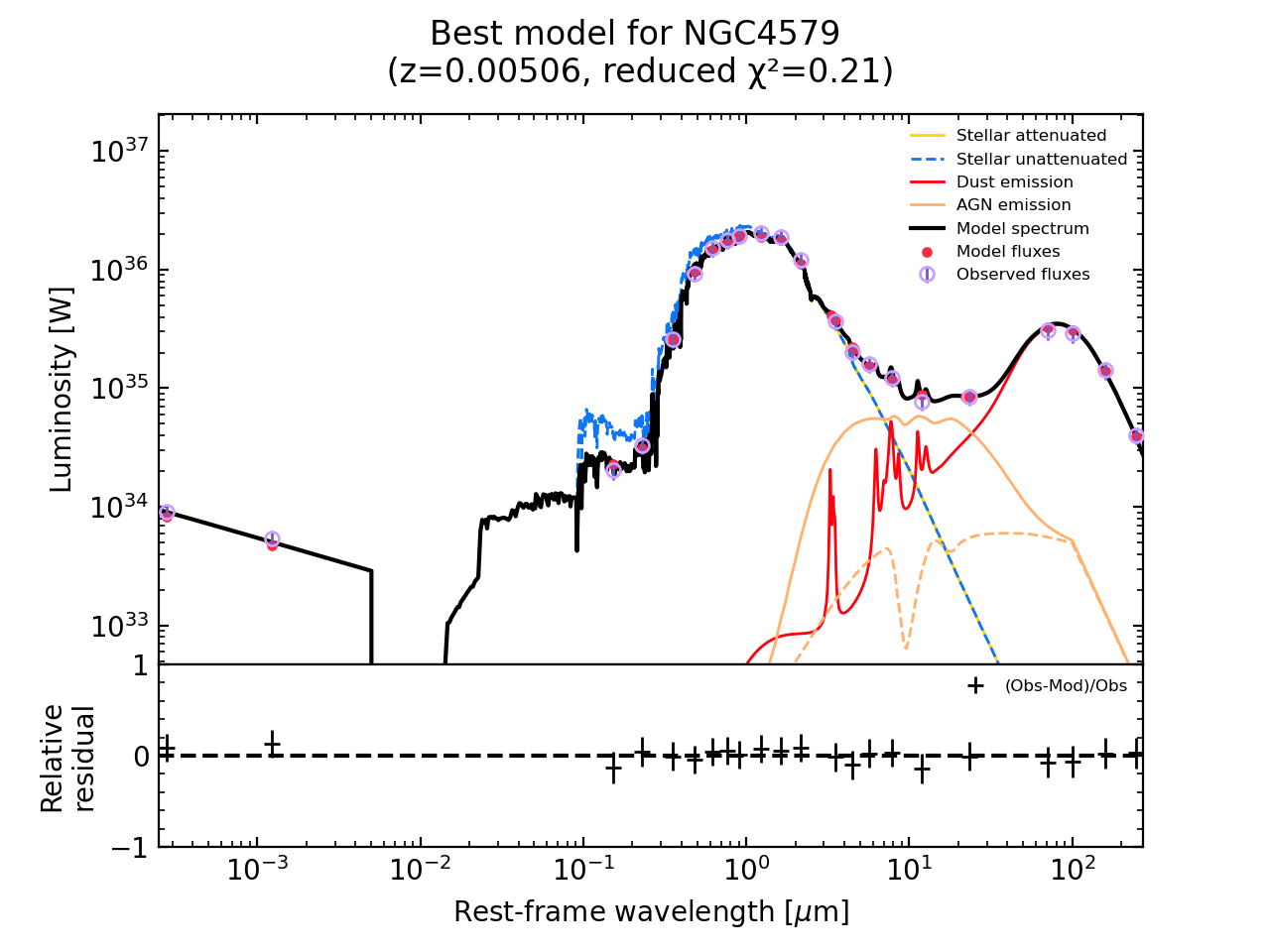}
    \includegraphics[width=0.49\linewidth]{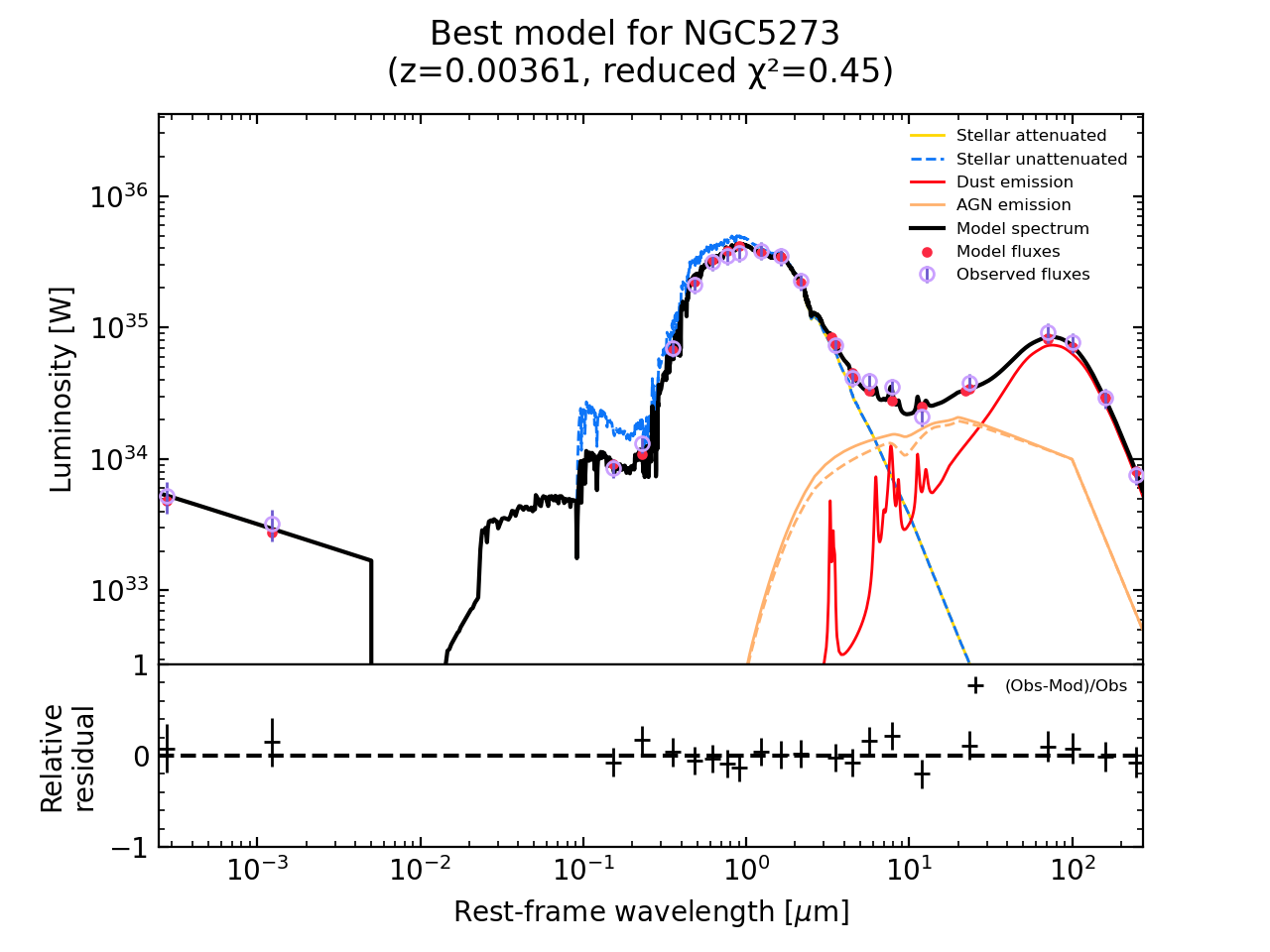}
    \caption[Example of SED fit]{Example of the best-fit achieved with our module for the LLAGN sample. Pink circles represent the observed photometry for each band, while red dots indicate the modeled flux in each band. The black line shows the best-fit composite model, with different colors representing specific model components: yellow denotes the stellar component attenuated by host galaxy extinction, the dashed blue line is the unattenuated stellar component, the brown line shows dust re-emission, and the solid orange line represents the total AGN emission. The dashed orange line indicates the attenuated central engine, thus the difference between the dashed and solid orange lines represents torus re-emission. NGC4036 and NGC4579 are presented on the left, showcasing a direct view of the central engine ($i=10^\circ$), with NGC4579 additionally exhibiting some dust extinction in the polar direction ($E(B-V)_\mathrm{polar} = 0.3$). Conversely, on the right, NGC1052 and NGC5273 are depicted, illustrating an obscured view of the central engine ($i=80^\circ$), highlighting the impact of viewing angle on the observed SED characteristics.}
    \label{fig:sed-fitting-examples}
\end{figure*}

The analysis of SED fitting results revealed a median reduced $\chi^2$ value of 1.00 and a mean value of 1.31, affirming consistently satisfactory fit quality. \Cref{fig:LLAGN-results} displays the distribution of essential AGN properties, including luminosities and Eddington ratios. The accretion luminosity ($L_\mathrm{Acc}$), representing the total radiation output of the central engine averaged across all directions, is contrasted with the bolometric luminosity ($L_\mathrm{Bol}$), which encompasses all AGN emissions, including dust re-emission. The measured luminosities for all sources span a broad range from $10^{39}$ to $10^{43}$~erg s$^{-1}$. Utilizing $L_\mathrm{Acc}$ alongside $M_\mathrm{BH}$ data sourced from the literature enabled the derivation of Eddington ratios predominantly within the range of $10^{-8}$ to $10^{-3}$, aligning with expectations for this sample. \Cref{fig:sed-fitting-examples} offers visual confirmations of the SED fitting outcomes for a selection of galaxies, showcasing the model’s proficiency in mirroring observed SED patterns. Additionally, a compilation of the primary physical parameters estimated from the CIGALE output is cataloged in \cref{table:06cigale-output}.

\begin{table*}[p]
    \centering
    \smaller
    \begin{tabular}{cccccccc}
    \hline\hline
    ID & $\log~L_\mathrm{Acc}$ & $\lambda_\mathrm{Edd}$  & $\log~L_\mathrm{Bol}$ & $\log~L_\mathrm{X}^\mathrm{mod}$ & $\log~L_\mathrm{2500\AA}$ &   $\log~L_\mathrm{2~keV}$ &  $\alpha_\mathrm{OX}$ \\
       & [erg/s]               &                         & [erg/s]               & [erg/s]                          &  [erg/s Hz]               & [erg/s Hz]                &  \\
    \hline
    NGC1052 & 42.63 $\pm$ 0.56 & 2.5E-04 &  42.18 $\pm$ 0.20 &  41.12 $\pm$ 0.11 & 25.67 $\pm$ 0.57 &  23.76 $\pm$ 0.74 & -0.73 $\pm$ 0.50 \\
    NGC2273 & 42.13 $\pm$ 0.61 & 6.7E-05 &  41.79 $\pm$ 0.29 &  40.87 $\pm$ 0.14 & 25.14 $\pm$ 0.59 &  22.94 $\pm$ 0.37 & -0.84 $\pm$ 0.36 \\
    NGC2685 & 39.80 $\pm$ 0.47 & 1.9E-06 &  39.31 $\pm$ 0.32 &  38.39 $\pm$ 0.22 & 22.82 $\pm$ 0.48 &  20.78 $\pm$ 0.43 & -0.78 $\pm$ 0.35 \\
    NGC2655 & 42.72 $\pm$ 0.62 & 1.7E-04 &  42.33 $\pm$ 0.14 &  41.20 $\pm$ 0.05 & 25.71 $\pm$ 0.62 &  24.03 $\pm$ 0.07 & -0.64 $\pm$ 0.26 \\
    NGC2768 & 40.62 $\pm$ 0.36 & 1.2E-06 &  40.05 $\pm$ 0.23 &  39.16 $\pm$ 0.14 & 23.64 $\pm$ 0.37 &  21.31 $\pm$ 0.31 & -0.89 $\pm$ 0.26 \\
    NGC2787 & 39.40 $\pm$ 0.60 & 7.2E-07 &  38.78 $\pm$ 0.32 &  37.88 $\pm$ 0.21 & 22.41 $\pm$ 0.60 &  21.47 $\pm$ 0.20 & -0.35 $\pm$ 0.31 \\
    NGC2841 & 39.33 $\pm$ 1.37 & 6.7E-08 &  39.20 $\pm$ 0.27 &  38.25 $\pm$ 0.18 & 22.33 $\pm$ 1.32 &  20.37 $\pm$ 0.43 & -0.75 $\pm$ 0.67 \\
    NGC3031 & 41.86 $\pm$ 0.46 & 4.8E-05 &  41.26 $\pm$ 0.16 &  40.41 $\pm$ 0.05 & 24.91 $\pm$ 0.46 &  22.11 $\pm$ 0.05 & -1.07 $\pm$ 0.19 \\
    NGC3079 & 40.69 $\pm$ 0.02 & 1.5E-04 &  40.53 $\pm$ 0.03 &  39.52 $\pm$ 0.05 & 23.70 $\pm$ 0.06 &  21.53 $\pm$ 0.22 & -0.83 $\pm$ 0.10 \\
    NGC3147 & 42.65 $\pm$ 0.95 & 1.7E-04 &  42.60 $\pm$ 0.17 &  41.82 $\pm$ 0.04 & 25.66 $\pm$ 0.91 &  23.53 $\pm$ 0.05 & -0.81 $\pm$ 0.37 \\
    NGC3185 & 41.14 $\pm$ 0.35 & 4.5E-05 &  40.51 $\pm$ 0.18 &  39.60 $\pm$ 0.09 & 24.15 $\pm$ 0.36 &  21.57 $\pm$ 0.20 & -0.99 $\pm$ 0.22 \\
    NGC3190 & 42.28 $\pm$ 0.71 & 6.9E-05 &  42.06 $\pm$ 0.27 &  41.20 $\pm$ 0.06 & 25.28 $\pm$ 0.70 &  25.19 $\pm$ 0.07 & -0.03 $\pm$ 0.30 \\
    NGC3193 & 40.33 $\pm$ 0.47 & 5.9E-07 &  39.80 $\pm$ 0.36 &  38.91 $\pm$ 0.25 & 23.34 $\pm$ 0.48 &  22.30 $\pm$ 0.25 & -0.40 $\pm$ 0.28 \\
    NGC3414 & 42.12 $\pm$ 0.46 & 6.5E-05 &  41.48 $\pm$ 0.20 &  40.35 $\pm$ 0.06 & 25.14 $\pm$ 0.46 &  22.43 $\pm$ 0.15 & -1.03 $\pm$ 0.23 \\
    NGC3516 & 42.92 $\pm$ 0.56 & 3.2E-04 &  42.50 $\pm$ 0.18 &  41.46 $\pm$ 0.10 & 25.92 $\pm$ 0.56 &  24.29 $\pm$ 0.23 & -0.62 $\pm$ 0.30 \\
    NGC3718 & 42.16 $\pm$ 0.65 & 2.0E-04 &  41.93 $\pm$ 0.14 &  41.04 $\pm$ 0.06 & 25.14 $\pm$ 0.65 &  22.80 $\pm$ 0.13 & -0.89 $\pm$ 0.30 \\
    NGC3898 & 39.02 $\pm$ 0.37 & 9.5E-08 &  38.94 $\pm$ 0.18 &  37.95 $\pm$ 0.12 & 22.04 $\pm$ 0.35 &  21.68 $\pm$ 0.26 & -0.13 $\pm$ 0.23 \\
    NGC3945 & 40.67 $\pm$ 0.96 & 3.2E-06 &  40.32 $\pm$ 0.33 &  39.34 $\pm$ 0.11 & 23.67 $\pm$ 0.94 &  22.73 $\pm$ 0.16 & -0.36 $\pm$ 0.42 \\
    NGC3953 & 40.00 $\pm$ 1.09 & 3.6E-06 &  39.58 $\pm$ 0.32 &  38.72 $\pm$ 0.14 & 23.00 $\pm$ 1.09 &  20.53 $\pm$ 0.21 & -0.94 $\pm$ 0.50 \\
    NGC3982 & 42.09 $\pm$ 0.50 & 2.0E-04 &  41.89 $\pm$ 0.23 &  41.02 $\pm$ 0.09 & 25.13 $\pm$ 0.54 &  23.50 $\pm$ 0.23 & -0.62 $\pm$ 0.29 \\
    NGC4036 & 42.14 $\pm$ 0.67 & 7.5E-05 &  41.71 $\pm$ 0.20 &  40.80 $\pm$ 0.11 & 25.15 $\pm$ 0.65 &  24.00 $\pm$ 0.25 & -0.43 $\pm$ 0.35 \\ 
    ... & ... &  ... &  ... &  ... &  ... &  ... & ... \\
    \hline
    \end{tabular}
    \caption[LLAGN Sample Properties]{Properties obtained from the SED fitting of the LLAGN sample. Luminosities in $\mathrm{erg~s^{-1}}$, except for monochromatic luminosities that are in $\mathrm{erg~s^{-1}~Hz^{-1}}$.}
    \label{table:06cigale-output}
\end{table*}

\section{Results and Discussion}
\label{IRX-sec:4}

This section delineates the outcomes derived from applying our novel CIGALE module, which is tailored to fit the SEDs of LLAGN. Specifically, \cref{subsec:4.1} validates our methodological approach, outlines the selection of our principal hypothesis, and contextualizes our findings within the existing literature. In \cref{subsec:4.2}, we explore the X-ray bolometric corrections within the low-luminosity regime. \Cref{subsec:4.3} studies the modeled $\alpha_\mathrm{ox}$ indices extracted from our SED fits, investigating their alignment with established scaling relations observed in QSOs. 

\subsection{Validation}
\label{subsec:4.1}

CIGALE incorporates a method to check how reliable its results are by performing a Bayesian analysis on its best-fit SED. This works by first creating a normal distribution for each filter, based on the best-fit SED's measured flux and its observational error. Then, random "mock" observations are created using these distributions, and the Bayesian analysis is done again. By comparing the known parameters from the original best-fit SED with the new parameters from this repeated analysis, the reliability of the results can be checked \citep[details in][]{Boquien19}.

\begin{figure*}[]
    \centering
    \includegraphics[width=0.99\linewidth]{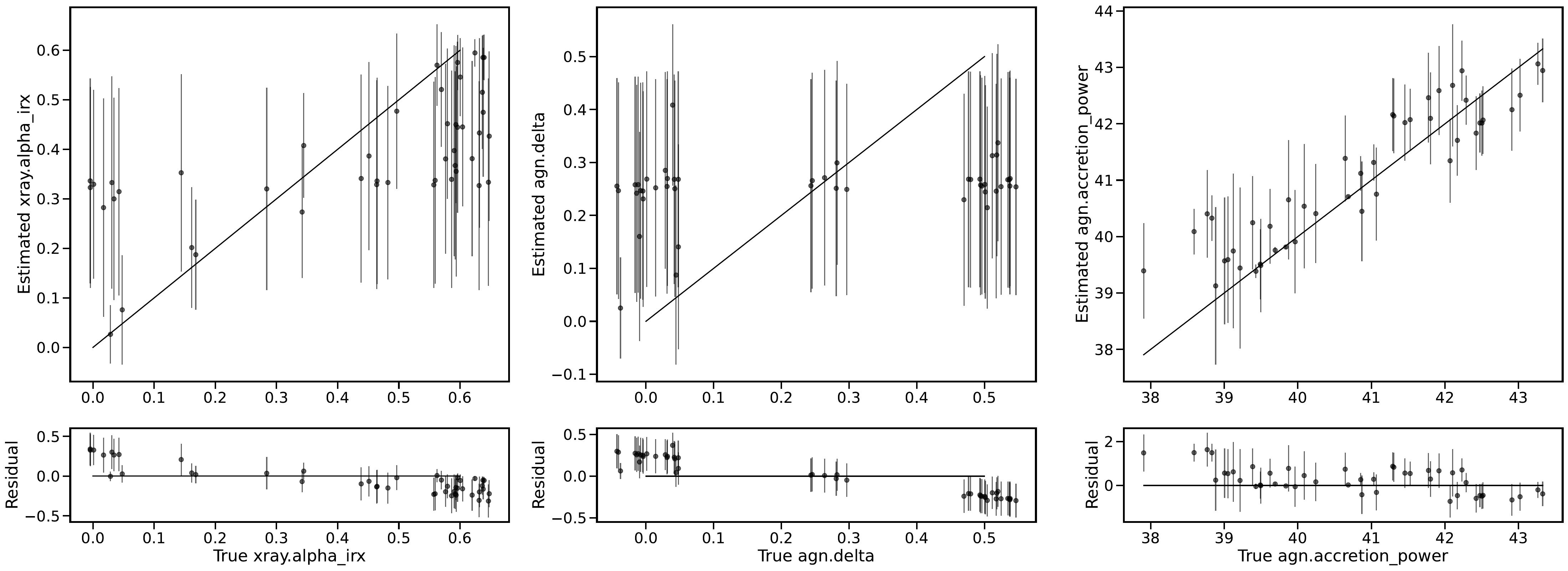}
    \caption[Validation]{The left panel shows the comparison between the true $\alpha_\mathrm{IRX}$ (from the best-fit SED) and the estimated values from the re-performed SED fitting. The solid black line represents the 1:1 relation, with residuals plotted at the bottom. The middle and right panels show similar comparisons for $\delta_\mathrm{AGN}$ and \accretion, respectively.}
    \label{fig:parameters}
\end{figure*}

We applied this method to our LLAGN sample, specifically focusing on the new parameters $\delta_\mathrm{AGN}$ and $\alpha_\mathrm{IRX}$, as well as the \accretion, which is the most important estimator for deriving bolometric luminosities from SED fitting. \Cref{fig:parameters} compares the "true" values of these parameters (derived from the best-fit SED) with the estimated parameters obtained from the re-performed SED fitting.

For most sources, $\delta_\mathrm{AGN}$ and $\alpha_\mathrm{IRX}$ are poorly constrained, though a few sources do show better constraints. This lack of precision is primarily due to the 9-arcsec aperture we used, which contaminates the MIR, the wavelength range most sensitive to these parameters, with host galaxy emission. Reducing the aperture size could increase the AGN fraction and lead to better constraints. However, the larger PSF of the far-IR bands imposes a limiting factor (for details, refer to \cref{appendix:aperture}). 

Importantly, in cases where $\delta_\mathrm{AGN}$ is poorly constrained, the current CIGALE module cannot reliably distinguish between different accretion disk states (e.g., pure ADAF, truncated disk). For sources where $\delta_\mathrm{AGN}$ exhibits large errors, meaning a broad range of values minimally affects the chi-square fit, it becomes challenging to specify the AGN emission's structural origin. This limitation should be considered when interpreting results related to the LLAGN central engine. To assess whether $\delta_\mathrm{AGN}$ is constrained for specific sources, we recommend running CIGALE with mock\_flag = True and analyzing the resulting outputs.

Despite these limitations on some AGN-specific parameters, their variation has minimal impact on the overall estimates of bolometric luminosities. While $\delta_\mathrm{AGN}$ is crucial for shaping the AGN’s UV-to-IR emission profile, it does not significantly affect the integrated luminosity values. Instead, the $L_\mathrm{X}$--$L_\mathrm{12\mu m}$ prior plays a central role in constraining both the accretion and bolometric luminosities by fixing the emission at 12 microns across all AGN models within a specified dispersion. This 12-micron luminosity constraint provides a stable reference point that limits the integral of the AGN emission, even if its shapes changes. As shown in \cref{fig:parameters}, most sources adhere to the 1:1 relation within the error margins, indicating that $L_\mathrm{bol}$ remains well-constrained, even when parameters like $\delta_\mathrm{AGN}$ are not tightly defined. The primary source of dispersion in the luminosity estimates results from variations in other parameters and their influence on the integrated emission.

\begin{figure}[h!]
    \centering
    \includegraphics[width=0.8\linewidth]{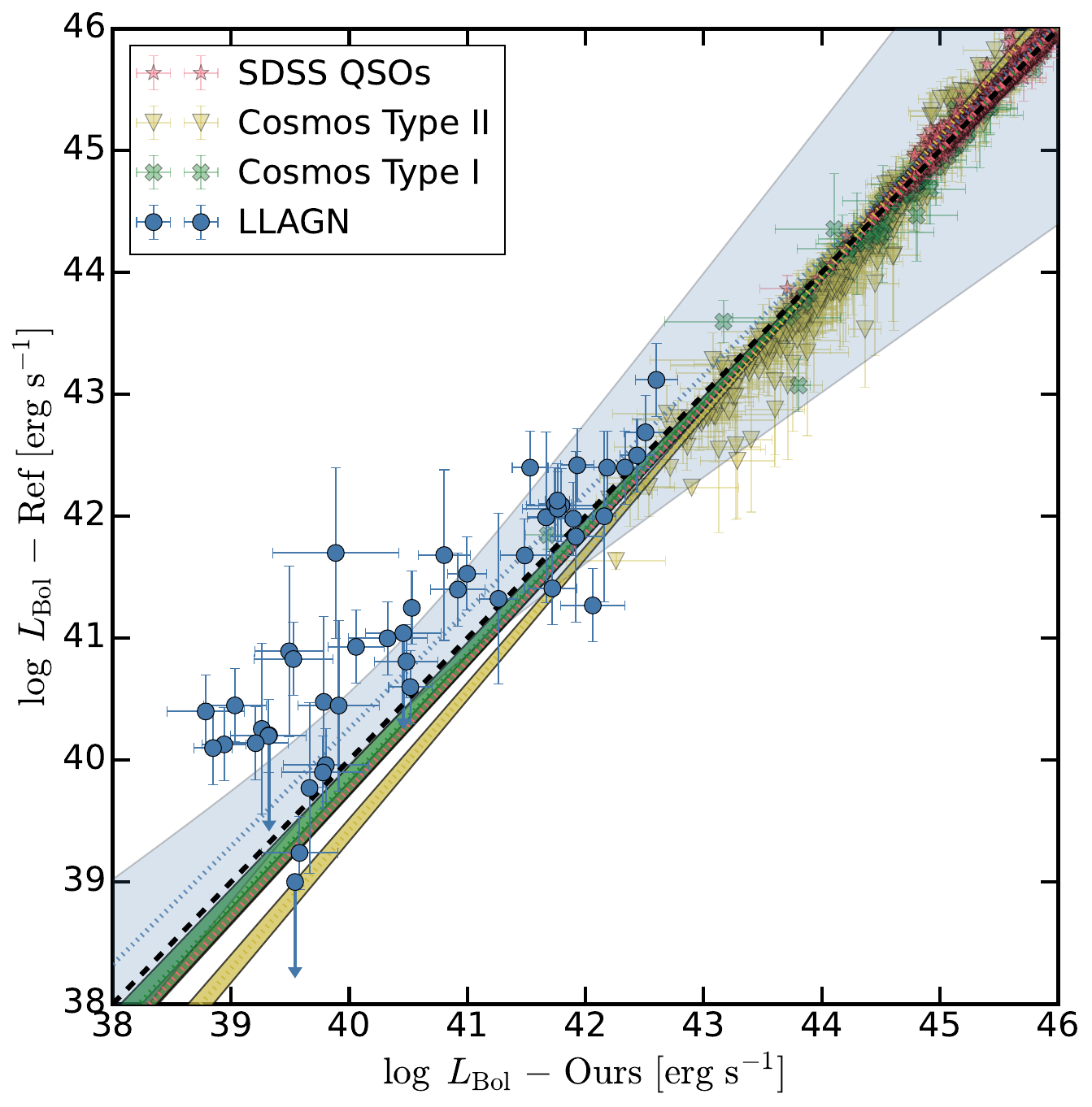}
    \caption[Validation]{Comparison of bolometric luminosities between this work and reference values. The dashed black line represents the 1:1 relation. A linear fit is shown for each sample with a dotted line following the same colors and with a shadow area of the 3-sigma deviation.}
    \label{fig:validation}
\end{figure}

As highlighted in \cref{subsec:2.1}, the $L_\mathrm{X}$--$L_\mathrm{12\mu m}$ relationship holds well in the low to mid-luminosity regime. While its extrapolation to high-luminosity sources can vary by an order of magnitude \citep{Stern2015ApJ...807..129S}, it remains reliable for $L_\mathrm{X} < 10^{45}\ \mathrm{erg\ s^{-1}}$ \citep{Asmus15}. We compared bolometric luminosities ($L_\mathrm{Bol}$) retrieved using the conventional CIGALE approach ($\alpha_{\mathrm{ox}}$) for the SDSS and COSMOS samples with those obtained using our new module. \Cref{fig:validation} compares these bolometric luminosities. The AGN samples show linear fits closely following the 1:1 relation, within error margins. However, for SDSS QSOs, a deviation from the 1:1 relation is observed at luminosities exceeding $10^{45}$ erg s$^{-1}$. 

In the low-luminosity regime, we adopted as reference value bolometric luminosities from \citet{Eracleous2010ApJS..187..135E} for the 14 common sources, and applied a bolometric correction of $k_\mathrm{X} = 15.8$ for the remaining sources \citep{Ho2009ApJ...699..626H}. For the LLAGN sample, a general luminosity trend is evident, although the reference luminosities are higher. This offset can be attributed to significant host galaxy contamination in the reference observations, which did not decompose the host galaxy from the AGN, as well as the use of a higher bolometric correction factor compared to the one we derived in \cref{subsec:4.2}.  Our luminosities are derived from AGN modeling using SED fitting, offering a more accurate representation of AGN luminosities by integrating the best-fit SED, which is especially crucial given the host galaxy's influence on AGN emissions. Moving forward, all future bolometric luminosities and Eddington values will be derived from our SED fitting module.

\begin{figure}[h!]
    \centering
    \includegraphics[width=1\linewidth]{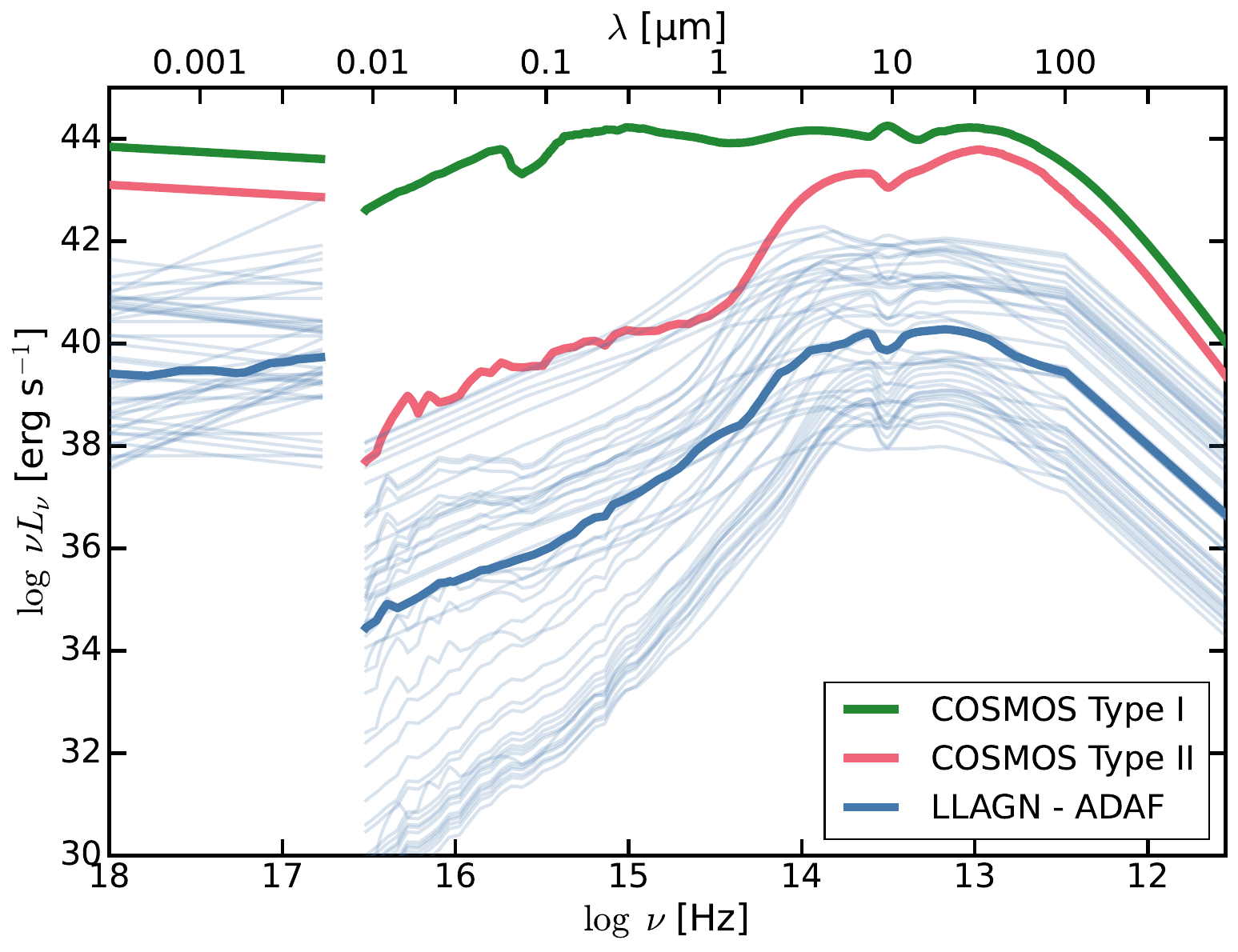}
    \caption[SED Comparison]{Comparison of the mean SEDs fitted for QSOs from COSMOS between Type I and Type II, alongside our mean LLAGN SED using the ADAF+TDk model. Transparent lines represent the individual SEDs of LLAGN that were fitted. X-ray emission is intrinsic and does not account for absorption.}
    \label{fig:mean-sed}
\end{figure}

In \cref{fig:mean-sed}, we compare the mean SEDs for the high-luminosity Type I and Type II AGN from the COSMOS sample with our LLAGN model. All SEDs include the central engine plus the toroid and polar dust emission. For the AGN sample, the conventional CIGALE approach is used; instead, the LLAGN model incorporates the new the central engine. As expected, the AGN SEDs reveal distinguishing characteristics between Type I and Type II AGN. Type II AGN exhibits weaker UV-optical emission due to the torus constraining the central engine. Beyond 1 micron and into the infrared bands, both SEDs become more similar, with a subtle divergence observed in the 9.7-micron silicate complex. This feature manifests as either absorption or emission depending on the optical depth at 9.7 microns, accompanied by a small bump emission at 50 microns for Type II AGN.

While QSOs encompass higher luminosities, both LLAGN models display comparatively lower luminosities. Initially, the mean SED of LLAGN shows characteristics of obscuration, as expected, since most LLAGN are Type II. This obscuration does not precisely match the shape of a Type II from COSMOS, exhibiting a less abrupt thermal emission at 1 micron. Transparent lines in \cref{fig:mean-sed} represent individual sources, revealing that some sources exhibit UV-optical slopes closer to Type II, indicative of unabsorbed ADAF. These non-obscured LLAGN showcase the expected UV shape (\cref{fig:ADAF-sed-model}). Introducing the ADAF model brings about a notable change in the infrared, where remission from polar dust and the toroid alters the power slope of the overall AGN emission.



\subsection{Bolometric correction on the lower-luminosity regime}
\label{subsec:4.2}

The X-ray bolometric correction ($k_\mathrm{X}$) is essential for converting observed X-ray luminosity ($L_\mathrm{X}$) of an AGN into its total bolometric luminosity. This correction addresses the fact that X-rays represent only a portion of the total radiation emitted by the AGN. However, X-rays play a significant role in conducting AGN surveys as AGN emissions dominate this energy band and can be easily identified. This becomes particularly vital in our LLAGN sample, where emissions from the host galaxy overshadow the UV-optical ranges, necessitating the consideration of the host galaxy's impact on bolometric luminosity.

SED fitting has proven to be an invaluable method for determining $L_\mathrm{Bol}$ of AGN. This technique comprehensively estimates total AGN luminosity as detailed in \cref{subsec:3.3}. By comparing $L_\mathrm{Bol}$ with $L_\mathrm{X}$, we derive the bolometric correction factor $k_\mathrm{X}$, such that $L_\mathrm{Bol}$\,$=$\,$k_\mathrm{X} \times L_\mathrm{X}$. The concept of X-ray bolometric correction has been the subject of detailed investigation in studies by \citet{Hopkins2007ApJ...654..731H}, \citet{Lusso12}, and \citet{Duras20}.  Notably, \citet{Duras20} developed a function for $k_\mathrm{X}$ that leverages SED fitting incorporating an infrared-X relation within a similar framework to ours, parametrizing how $k_\mathrm{X}$ varies with luminosity:

\begin{equation}
    k_{\mathrm{X}}\left(L_{\mathrm{Bol}}\right)=a\left[1+\left(\frac{\log \left(L_{\mathrm{Bol}}/\mathrm{[erg\,s^{-1}]}\right)}{b}\right)^c\right] .
    \label{eq:bolbol}
\end{equation}

This formula demonstrates the dependence of $k_\mathrm{X}$ on bolometric luminosity, which becomes flatter below a specific luminosity threshold. Extrapolations from \citet{Duras20} suggest a $k_\mathrm{X}$\,$=$\,$10.96$ for the lower luminosity regime, while \citet{Eracleous2010ApJS..187..135E} and \citet{Ho2009ApJ...699..626H} propose $k_\mathrm{X}$\,$=$\,$50$ and $k_\mathrm{X}$\,$=$\,$15$, respectively.


In \cref{fig:kbol-correction}, we compare the relationship between $L_\mathrm{Bol}$ and $L_\mathrm{X}$ for our LLAGN sample, with corrections from \citet{Nemmen14}, \citet{Eracleous2010ApJS..187..135E}, \citet{Ho2009ApJ...699..626H} and the low-luminosity extrapolation from \citet{Duras20}. The Pearson correlation coefficient is approximately 0.9. A linear regression performed on our data yields:

\begin{figure}[!t]
    \centering
    \includegraphics[width=1\linewidth]{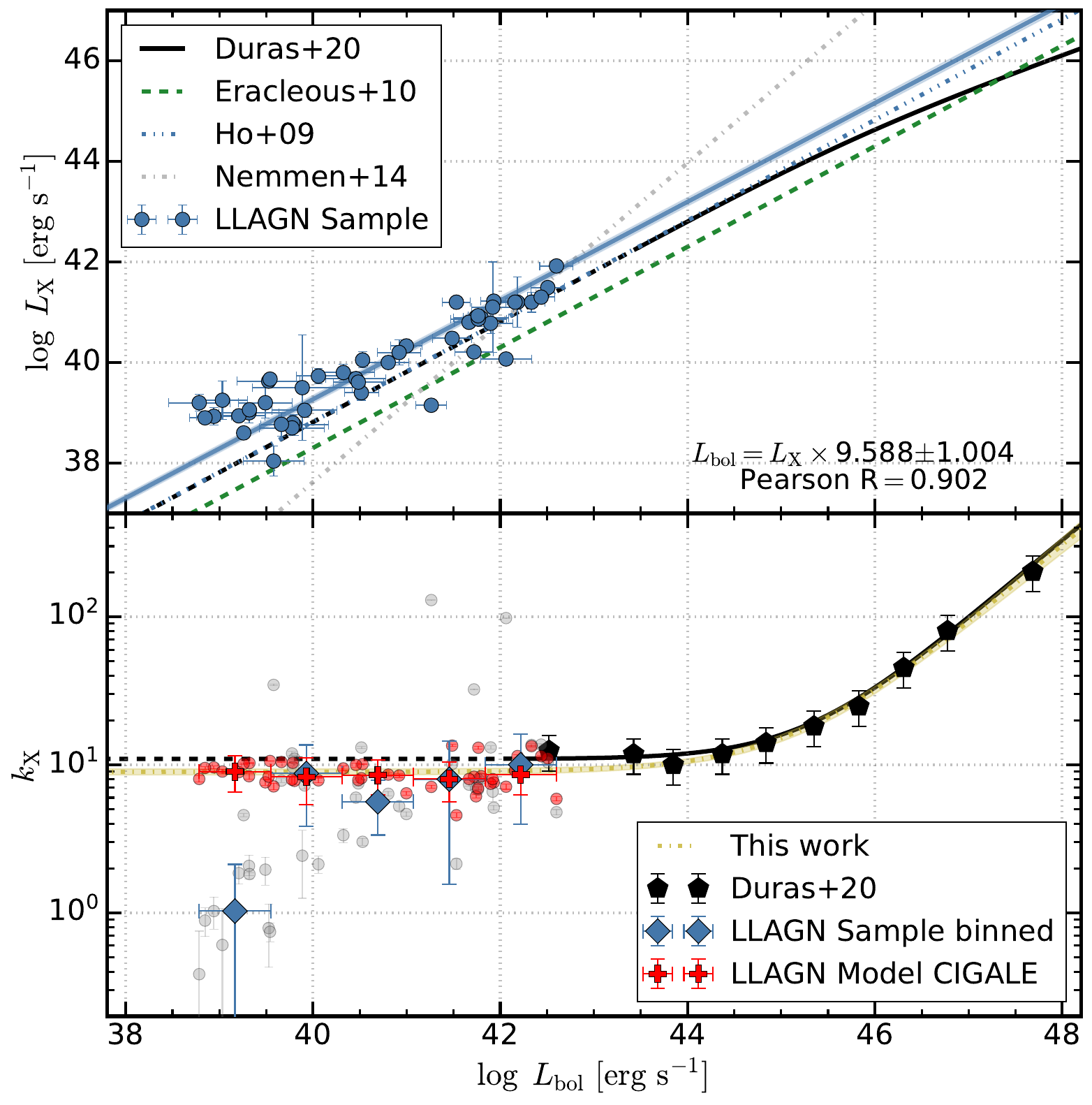}
    \caption[Bolometric correction]{Upper panel: Blue dots depict the measured X-ray luminosities plotted against the bolometric luminosities for our LLAGN sample, with a blue line indicating the linear regression fitted to this data. In contrast, the data and bolometric correction from \citet{Nemmen14} are shown as grey stars and a dotted line, respectively. The solid black line represents the bolometric correction from \citet{Duras20}, transitioning to a dotted line where it extrapolates beyond the data. Lower panel: X-ray bolometric correction ($k_\mathrm{X}$) plotted against bolometric luminosity. The solid black line denotes the bolometric correction from \citet{Duras20}, accompanied by black pentagons representing its binned data. Our binned data for the LLAGN sample are illustrated with blue diamonds. The dotted yellow line represents the fit to \cref{eq:bolbol} using a combination of our binned data and that from \citet{Duras20}. Red crosses mark the binned data from the pure AGN X-ray luminosities modeled by CIGALE. Shaded regions indicate the 1-sigma uncertainties.}
    \label{fig:kbol-correction}
\end{figure}

\begin{equation}
    L_\mathrm{Bol} = (9.588 \pm 1.004) \times L_\mathrm{X}.
    \label{eq:bol}
\end{equation}

This fit indicates a $k_\mathrm{X}$ value lower than those typically reported in the literature, yet it aligns closely, within error margins, with the extrapolation by \citet{Duras20}. Differences from previous studies can be attributed to several factors in our methodology: our specific modeling of the accretion process, a larger sample size that reduces the impact of statistical fluctuations, and the inclusion of host galaxy X-ray emissions in our modeling.



As illustrated in the lower panel of \cref{fig:kbol-correction}, we present our $k_\mathrm{X}$ values for all LLAGN sources, taking into account the observed $L_\mathrm{X}$ (both raw and binned), the binned $L_\mathrm{X}$ from the AGN modeled by CIGALE, and the binned data from \citet{Duras20} for comparison in the high-luminosity regime. Additionally, we include a fit using \cref{eq:bolbol}, considering our modeled $k_\mathrm{X}$ (the ratio of bolometric luminosity to X-ray luminosity modeled by CIGALE), and incorporating the data from Duras. This approach enables us to provide an expression for $k_\mathrm{X}$ that spans nearly ten orders of magnitude in luminosity\footnote{\citet{Duras20} use an X-ray-IR relation as a prior in their SED fitting, facilitating a strong comparison.}. The fit achieves a reduced $\chi^2$ of 0.2, with parameters $a = 8.97 \pm 0.40$, $b = 45.28 \pm 0.11$, and $c = 60.00 \pm 3.40$.

We also plot the binned data based on observational $L_\mathrm{X}$ and $L_\mathrm{Bol}$ derived from SED fitting rather than using the AGN modeled $L_\mathrm{X}$. This reveals a lower $k_\mathrm{X}$ value in the first bin, suggesting that the host galaxy's X-ray emission becomes comparable to that of the AGN, affecting measurements on these luminosities.

\begin{figure}[]
    \centering
    \includegraphics[width=1\linewidth]{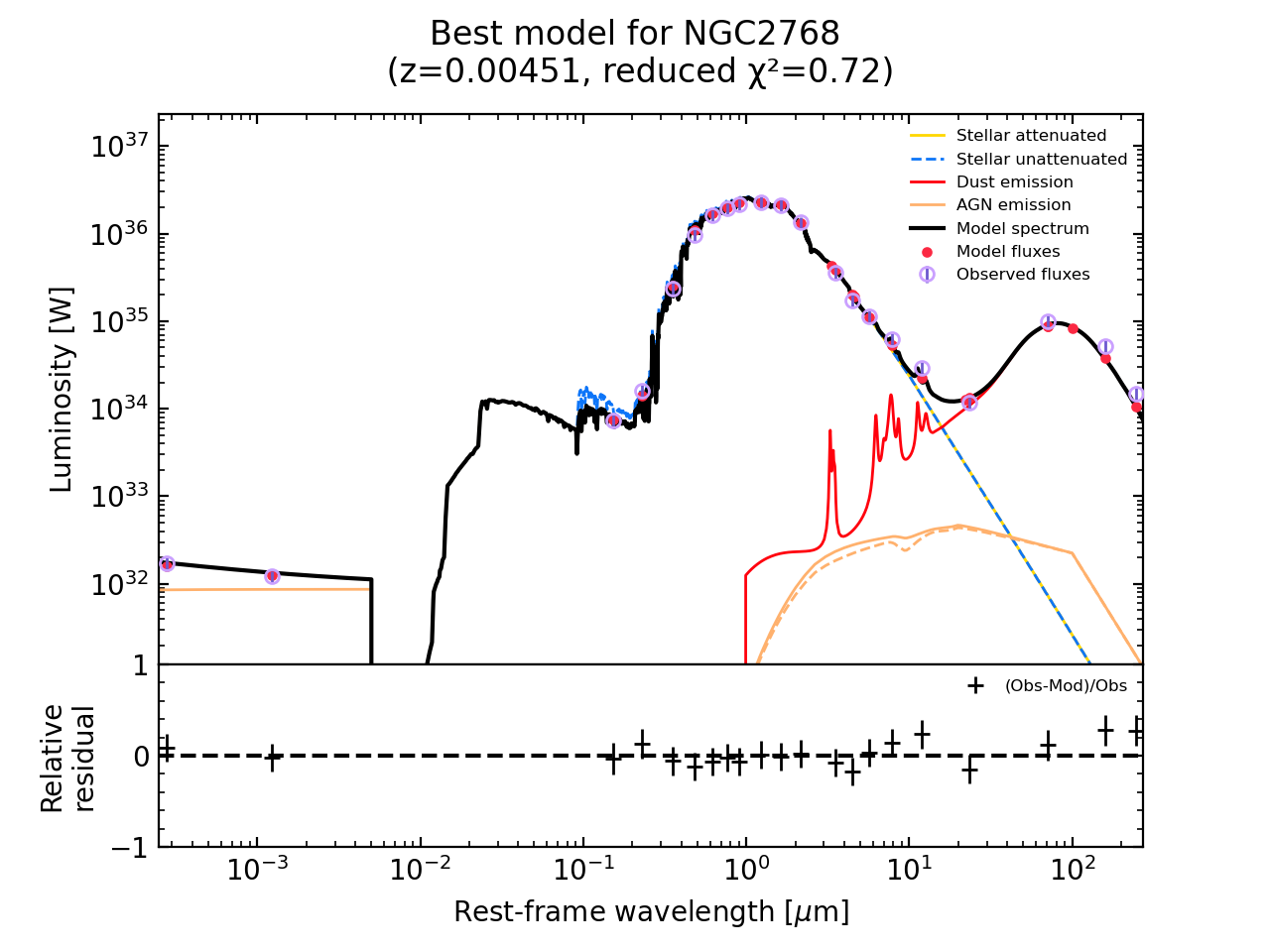}
    \caption[Example of SED fitting for the LLAGN sample]{Example of SED fitting for the LLAGN sample, where the host galaxy can emit in the X-ray hard band, modifying the contribution of an AGN to this band.}
    \label{fig:sed-fitting-xray}
\end{figure}

The key difference between observed X-ray luminosities and those modeled solely for AGN by CIGALE is that CIGALE assumes the host galaxy's X-ray emission primarily originates from low-mass X-ray binaries (LMXBs), high-mass X-ray binaries (HMXBs), and hot gas \citep{Yang20}. The CIGALE X-ray module applies the recipe from \citet{Mezcua2018MNRAS.478.2576M} to model the X-ray emission from these components, with LMXB and HMXB luminosities modeled as functions of stellar age and metallicity. Hot gas emission, on the other hand, is quantified in terms of the SFR. While shocks, such as those from supernovae or AGN feedback, can also produce X-ray emissions, CIGALE does not yet include models for shock-generated X-rays. \Cref{fig:sed-fitting-xray}  highlights a source as an example where the hard X-ray band significantly contributes to the host galaxy's overall X-ray emission, demonstrating CIGALE's capability to differentiate between contributions from LLAGN and the host galaxy, thereby improving the accuracy of the bolometric correction.

The debate surrounding the accretion mechanisms in LLAGN 
motivated the inclusion of an ADAF and a possible truncated thin disk, as elaborated in \cref{subsec:2.2}. Although it is hypothesized that all sources in our sample undergo a non-standard form of accretion, only 60\% have independent evidence supporting this premise. To assess the potential for a traditional accretion mechanism, we forced a SED fitting with a standard accretion disk ($\delta_\mathrm{AGN} = 1$). Analysis of the reduced chi-square values revealed comparable fitting quality between the ADAF and conventional disk accretion models, highlighting a limitation in employing SED fitting as a discriminant between these models. Moreover, the bolometric luminosities derived from both ADAF and disk models are within error margins of each other. It is noteworthy, however, that bolometric luminosities obtained via the disk model are marginally higher, exhibiting an average difference of 0.1 dex. This finding indicates a notable consistency in the X-ray bolometric correction across different accretion scenarios.


This comparison reveals that although the distribution of UV-to-IR photons may vary, the total integrated luminosity exhibits no significant dependency on the chosen accretion model. This independence arises because the $L_\mathrm{X}$--$L_\mathrm{12\mu m}$ relationship, which serves as a foundational prior for AGN emission, effectively balances the SED profile. This equilibrium compensates for the lack of UV photons with an increased presence of IR emissions, resulting in a bolometric correction that remains unaffected by the accretion mode. This finding underscores the model-independence of bolometric corrections, emphasizing their reliability across different accretion regimes.

\subsection{Galaxy contamination}
\label{subsec:4.3}

Exploring the influence of different photometric apertures on the SED fitting for our LLAGN sample sheds light on how solid this module is for future applications. We specifically evaluated how different the AGN luminosity obtained from our 9-arcsec central photometry is from those obtained from entire galaxy photometry provided by DustPedia.

\begin{figure}[!t]
    \centering
    \includegraphics[width=0.8\linewidth]{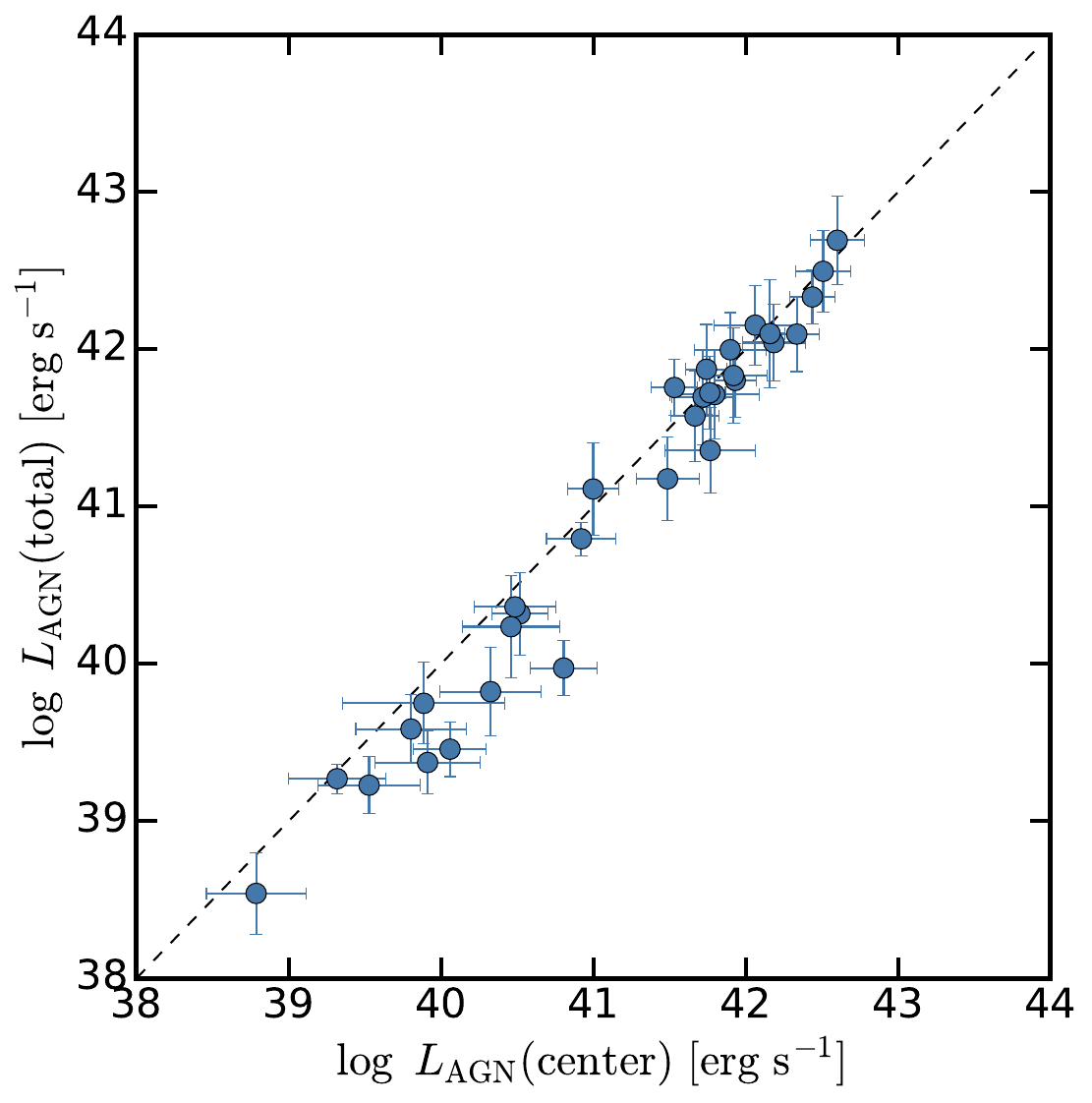}
    \caption[Bolometric luminosities in case of galaxy contamination]{Comparison of bolometric luminosities obtained using a central aperture-based photometric approach and entire galaxy photometry for the LLAGN sample. The dotted line represents the 1:1 relationship. Excluding a few outliers, our sources generally exhibit lower luminosities for the entire galaxy fit but align with the 1:1 relation within uncertainties. This suggests that bolometric luminosity can be accurately estimated even in scenarios with significant host galaxy contamination.}
    \label{fig:full-galaxy}
\end{figure}

\Cref{fig:full-galaxy} compares bolometric luminosities calculated using these two distinct methods. The congruence observed in the results indicates that our module effectively captures the bolometric luminosities of LLAGN across a wide range of luminosities. Additionally, the findings from the aperture-based method closely align with those obtained using full galaxy photometry. While the full galaxy photometry tends to show slightly lower luminosities than the central aperture, it still follows the 1:1 relation within the uncertainties, indicating that the discrepancies are minimal and fall within the error margins. The average offset between the two methods is 0.16 dex.

This consistency in bolometric luminosity estimation becomes particularly pertinent for sources at higher redshifts. At such distances, an instrument's PSF might encapsulate emissions from the central regions and the periphery of galaxies, potentially complicating the task of distinctly resolving galactic structures. Despite these challenges, the IRX module demonstrates a steadfast capability to accurately estimate bolometric luminosities, demonstrating its effectiveness even when full galaxy photometry is employed. These observations highlight the utility of this module when the photometry accounts for the galaxy's total emitted flux, extending its applicability to high-redshift observations.

\subsection{$\alpha_{OX}$ in Low-Luminosity AGN}
\label{subsec:4.4}

Our approach does not dictate a specific value for $\alpha_\mathrm{ox}$ but instead facilitates the reconstruction of the modeled 2500,\AA\ emission at its core to compare with the 2\,keV emission. This methodology allows for examining the $\alpha_\mathrm{ox}$ behavior within the modeled framework, offering predictions for the low-luminosity domain.

\begin{figure}[h!]
    \centering
    \includegraphics[width=0.8\linewidth]{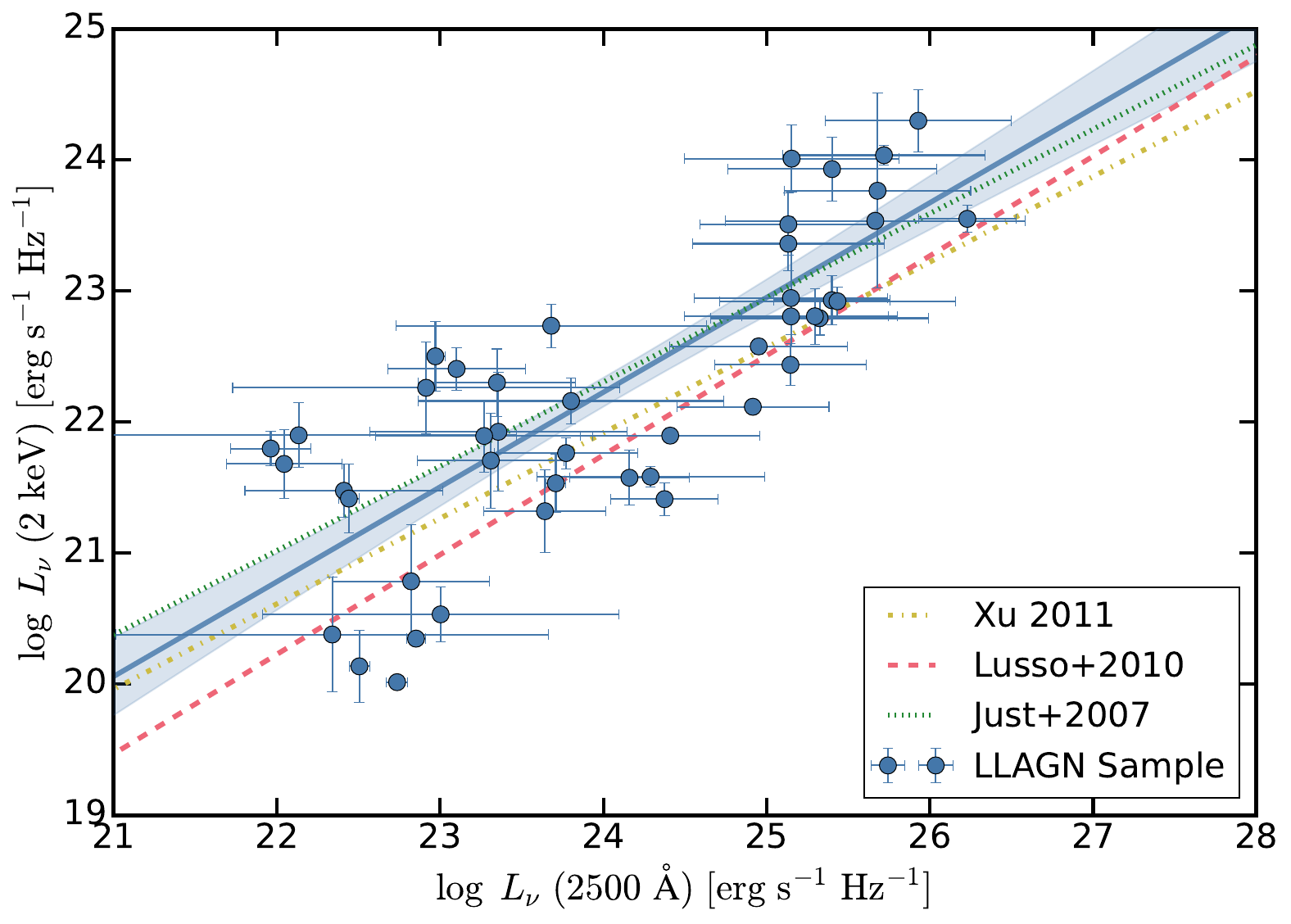}
    \caption[Comparison between UV and X-ray luminosities]{Comparison of the relationship between 2 keV and 2500\,\AA\ emissions with \citet{Lusso10}, \citet{Xu11}, and \citet{Just07} data. Our linear fit is a solid blue line, with shaded regions being the uncertainties at 1-$\sigma$.}
    \label{fig:2kev_vs_2500}
\end{figure}

The fundamental premise of the $\alpha_\mathrm{ox}$--$L_\mathrm{UV}$ relationship is the interplay between 2\,keV and 2500\,\AA\ emissions. Figure \ref{fig:2kev_vs_2500} illustrates this relationship and compares it with the linear correlations observed in both high- and low-luminosity regimes, as reported by studies such as \citet{Lusso10}, \citet{Just07}, and \citet{Xu11}. Our analysis shows a shift towards lower UV luminosities at comparable X-ray luminosities. Although a linear fit approximates the overall trend, the residuals reveal significant scatter ($\sigma$\,$\sim$\,$0.72$). The equation for the linear fit to our data is as follows:

\begin{equation}
    \log~L_\nu\mathrm{(2~keV)}=(0.72 \pm 0.09) ~\times~ \log~L_\nu\mathrm{(2500\AA)}+(4.88 \pm 2.13)
\end{equation}

\begin{figure}[t!]
    \centering    
    \includegraphics[width=0.8\linewidth]{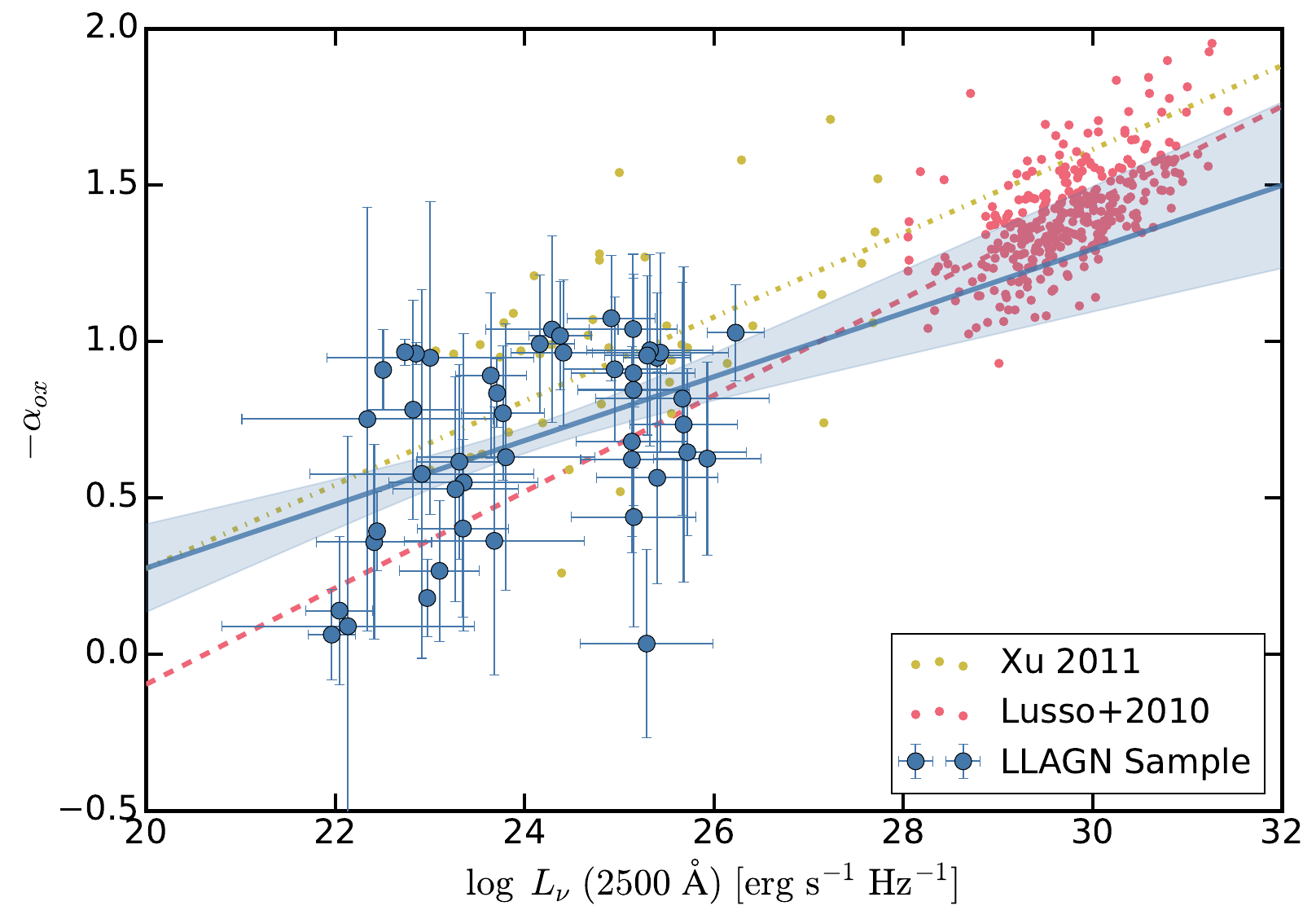}\\
    \includegraphics[width=0.8\linewidth]{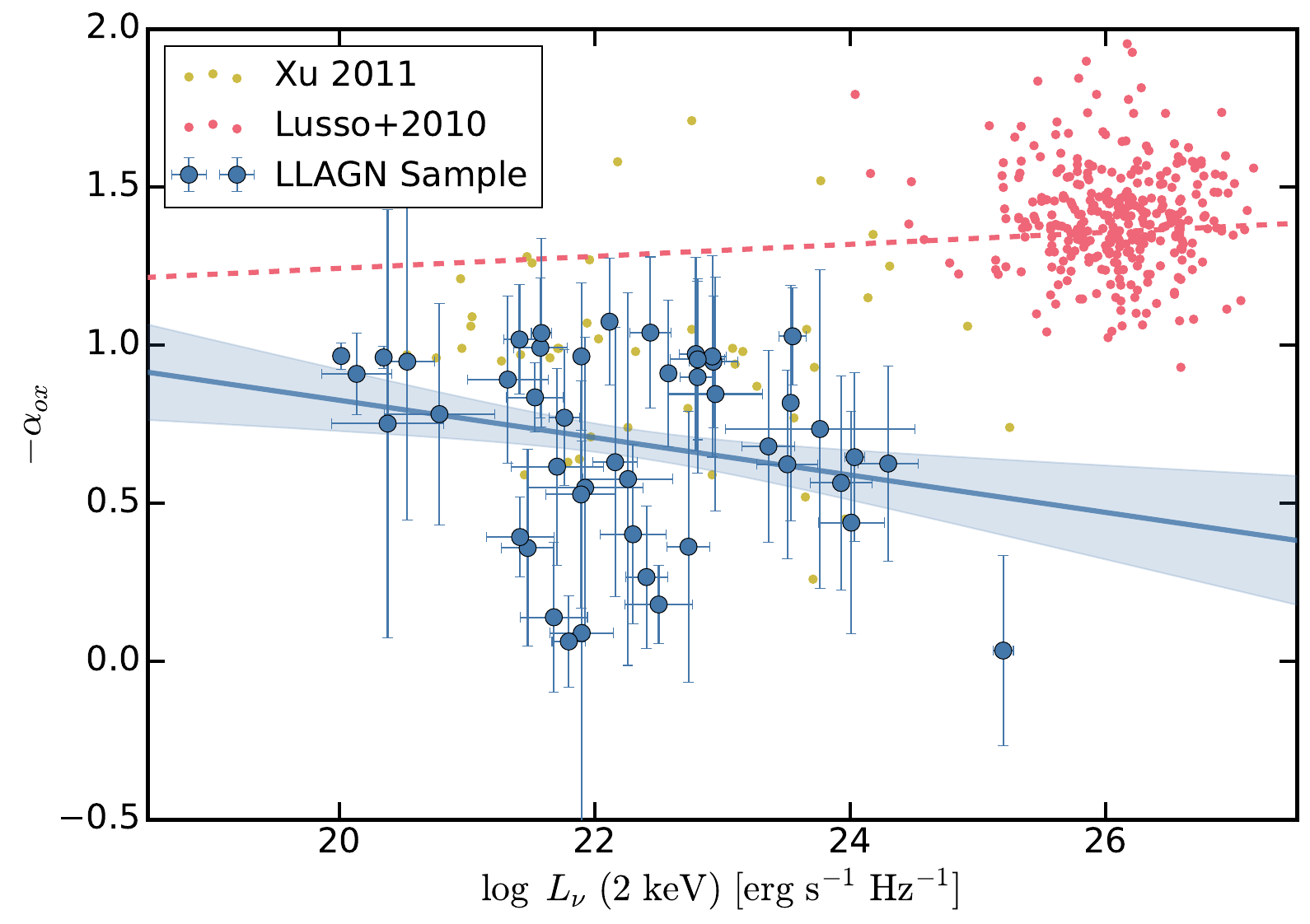}
    \caption[Comparison of $\alpha_\mathrm{ox}$ between AGN and LLAGN]{Comparison of spectral index $\alpha_\mathrm{ox}$ with 2500\,\AA\ luminosity (top) and two keV X-ray luminosity (bottom). The AGN sample by \citet{Lusso10} is in red, the LLAGN sample by \citet{Xu11} is in yellow, and our values from the model LLAGN for each source are in blue.}
    \label{fig:alpha_vs_lum}
\end{figure}

Figure \ref{fig:alpha_vs_lum} presents the $\alpha_\mathrm{ox}$ relationship with UV and X-ray luminosities. In the QSO regime, a pronounced correlation exists between $\alpha_\mathrm{ox}$ and UV luminosity, which is less apparent with X-ray luminosity. \citet{Lusso10} found $\alpha_\mathrm{ox}$ to loosely correlate with $L_\mathrm{2500\AA}$, exhibiting more scatter with $L_\mathrm{2keV}$. \citet{Xu11} noted a similar pattern for $L_\mathrm{2500\AA}$ within the LLAGN regime, albeit with greater dispersion, while no distinct correlation emerged for $L_\mathrm{2keV}$. In the QSO regime, the predominance of UV photons, primarily from the big blue bump, over corona-generated X-rays domain the relation and determines the $\alpha_\mathrm{ox}$.

Our findings in the lower luminosity range, as illustrated in blue in \cref{fig:alpha_vs_lum}, uncover a weak correlation of $\alpha_\mathrm{ox}$ with UV luminosity, characterized by a slope of $0.10 \pm 0.03$, and a slight inverse correlation with X-ray luminosity, indicated by a slope $-0.06 \pm 0.04$. The dispersion of residuals in both cases is $\sigma$\,$\sim$\,$0.27$. The $\alpha_\mathrm{ox}$ values in LLAGN primarily range between 0 and -1, covering four X-ray and UV luminosities magnitudes. Furthermore, our data straddle the extrapolated fit from \citet{Lusso10} in the 2500\,\AA\ luminosity plane, suggesting that neither consistent nor variable behaviors can be conclusively dismissed, especially considering the margin of error.

While UV and X-rays exhibit correlation in the high-luminosity regime, the source of X-ray photons in low-luminosity AGN may not solely be a hot corona. ADAFs are known for their pronounced emission in hard X-rays. Hence, X-ray photons in this context could stem from various processes, including inverse-Compton scattering, but operating at disparate scales. This complexity could account for the observed significant dispersion in the $L_\mathrm{2keV}$--$L_\mathrm{2500\AA}$ relationship.

We further examine the $\alpha_\mathrm{ox}$--$\lambda_\mathrm{Edd}$ domain to gain deeper insights. The correlation found in QSOs sources by \citet{Lusso10}, where sources with higher accretion rates showcase elevated $\alpha_\mathrm{ox}$ values, corroborates the notion that increased accretion intensifies the big blue bump, thereby boosting UV photon production. In contrast, an anticorrelation in LLAGN was observed by \citet{Xu11}, suggesting a divergent behavior in this regime.

\begin{figure}
    \centering
    \includegraphics[width=1\linewidth]{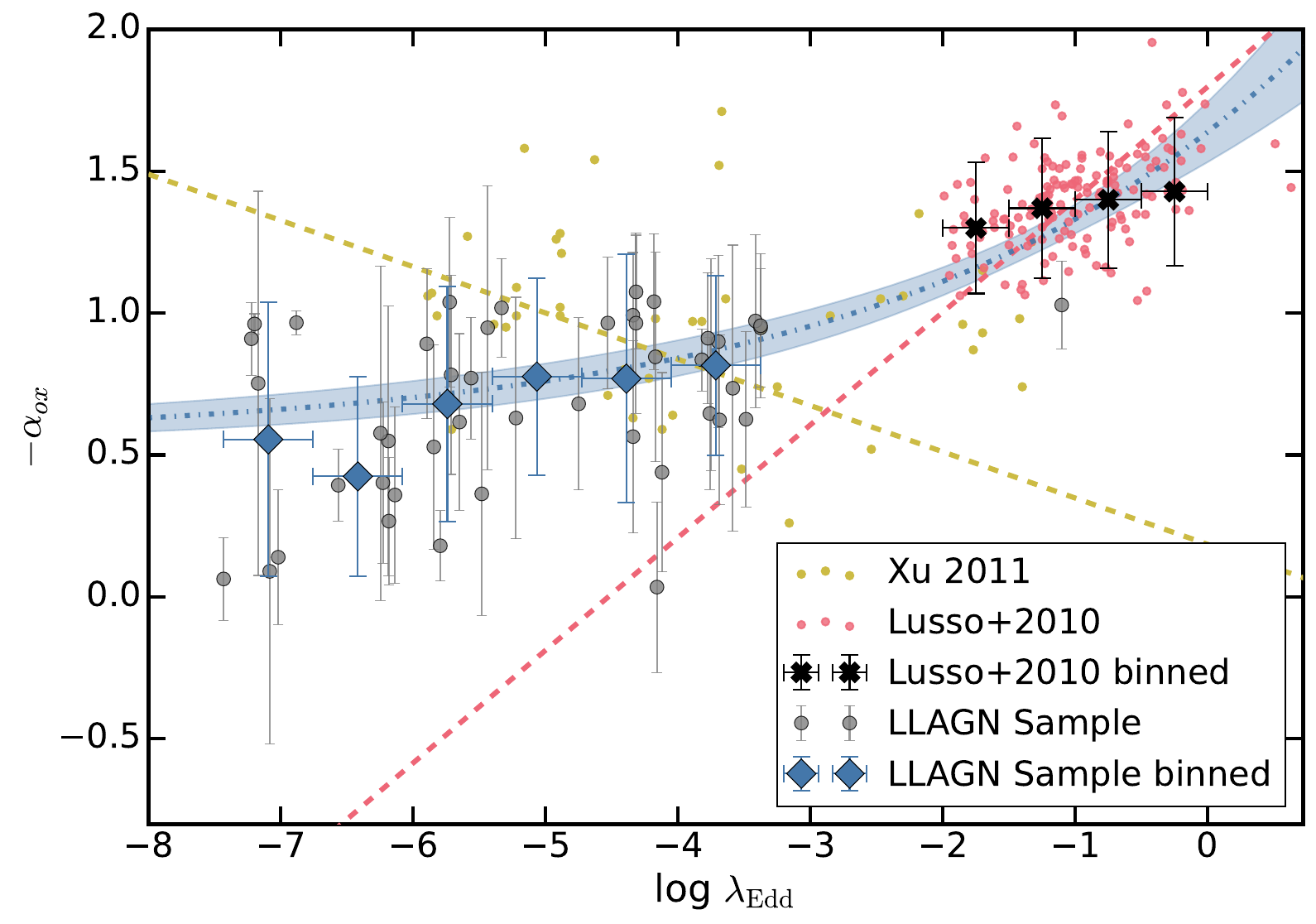}
    \caption[Comparison of $\alpha_\mathrm{ox}$ and $\lambda_\mathrm{Edd}$ between AGN and LLAGN]{Comparison of the spectral index $\alpha_\mathrm{ox}$ with the Eddington ratio ($\lambda_\mathrm{Edd}$). The AGN sample from \citet{Lusso10} is represented in red, the LLAGN sample from \citet{Xu11} is in yellow, and our modeled values for each LLAGN source are in grey. Our binned data are depicted as blue diamonds, while the black crosses represent the binned data from \citet{Lusso10}. The dashed blue line indicates the fit to all the binned data. Contrary to the anticorrelation proposed by \citet{Xu11}, our fit follows a constant value for the low regime. Shaded regions indicate the 1-sigma uncertainties.}
    \label{fig:alpha_vs_lambda}
\end{figure}

Our findings, shown in \cref{fig:alpha_vs_lambda}, indicate that LLAGN tend to display either a weak correlation or maintain constant $\alpha_\mathrm{ox}$ values across different accretion rates. The average $\alpha_\mathrm{ox}$ value for LLAGN is -0.69, compared to 1.4 for \citet{Lusso10} sample, highlighting a distinct difference between the two populations. By binning our data, combining it with the dataset from \citet{Lusso10}, and applying a broken power-law fit, we capture the behavior across nine orders of magnitude in $\lambda_\mathrm{Edd}$.

\begin{equation}
    -\alpha_\mathrm{ox}\left(\lambda_\mathrm{Edd}\right)=a\left[1+\left(\frac{\log \left(\lambda_\mathrm{Edd}\right)}{b}\right)^c\right] .
    \label{eq:acc}
\end{equation}

Since the accretion regime is supossed to switch at $\log~\lambda_\mathrm{EDD}$\,$=$\,$-2$, we anchor $b$\,$=$\,$0.01$. The fit yields a $\chi^2_\mathrm{red}$ of 1.39, characterized by parameters $a=0.55 \pm 0.02$ and $c=0.14 \pm 0.02$. \Cref{fig:alpha_vs_lambda} showcases the behavior of the relation. 

This outcome is consistent with the UV/X-ray correlation depicted in \cref{fig:2kev_vs_2500}, suggesting a stable ratio between these emissions and a uniform $\alpha_\mathrm{ox}$. Whereas in conventional QSOs, variations in accretion rates can modify the big blue bump across the UV-optical spectrum, in the low-luminosity regime, the accretion flows predominantly generate the optical/UV photons. This process involves the inverse Compton scattering of soft synchrotron photons by hot electrons within the ADAF, a phenomenon governed more by magnetic field dynamics than by mass influx into the flow.

Nonetheless, a deeper exploration is warranted to fully grasp this accretion paradigm's ramifications. Measuring UV emissions in LLAGN is complicated by their intrinsic faint nature, and our analysis leverages ADAF-centric modeling. For instance, the study by \citet{Xu11} relies on UV luminosity estimations derived from optical photon proxies, like continuum or emission lines. Future research utilizing high-resolution data in the UV, NIR, and MIR from current instruments like the \textit{Hubble Space Telescope} (HST), \textit{Euclid} and \textit{James Webb Space Telescope} (JWST), will be essential for more precisely delineating the SED in the central parsec of these galaxies. Additionally, next-generation FIR instruments with improved angular resolution will constrain better the FIR emission, further refining our understanding of LLAGN accretion processes.

\section{Summary and Conclusions}
\label{IRX-sec:5}

This work introduces a novel CIGALE module specifically designed for the analysis of low-luminosity AGN. By integrating empirical relationships and physically-based accretion models, this module improves our understanding of LLAGN through SED fitting, without relying on standard QSO extrapolation. Key findings from our comparison of local LLAGN properties to those of QSOs using this new module include:

\begin{enumerate}
    \item We developed a new X-ray CIGALE module that leverages the empirical $L_\mathrm{X}$--$L_\mathrm{12\mu m}$ relationship to link AGN X-ray and infrared emissions. This provides a new prior for estimating AGN emission across UV-to-IR bands, applicable to LINERs and Seyferts up to $10^{45}$ erg/s, and successfully tested in LLAGN.
    \item We incorporated a new seed photon model for AGN that uses a single parameter to transition from a pure ADAF spectrum to a standard accretion disk spectrum (see \cref{subsec:2.2.1}). This also accounts for intermediate states where the ADAF dominates inner orbits, while a truncated disk governs the outer orbits, resulting in a mix of non-thermal and thermal emissions. This physically-motivated approach introduces greater diversity in the templates that CIGALE can now produce (see \cref{fig:mock_seds}).
    \item Both additions were tested using mock catalogs, showing good retrieval of the true parameters, though with a slight bias toward higher input values. The recovery of AGN accretion power was tested across a range of AGN fractions, with most of the distribution of recovered minus true values falling within 0.5 dex (for details, see \cref{subsec:2.3}).
    \item We compiled a dataset of 50 X-ray-detected local galaxies, comprising LINERs and Seyferts, to study the lower luminosity range. We compared these luminosities with previous work and tested the recovery of input parameters, revealing that 9-arcsecond photometry is insufficient for constraining these parameters, though the AGN luminosity was successfully constrained. A mid-luminosity sample of AGN from COSMOS and SDSS was also used to test the implementation of the $L_\mathrm{X}$--$L_\mathrm{12\mu m}$ relation, yielding bolometric luminosities comparable to those retrieved by the previous X-ray CIGALE module from \citet{Yang20}.
    \item The new module demonstrated accuracy in estimating LLAGN bolometric luminosities, showing a small 0.18 dex difference when using full-galaxy photometry, effectively minimizing galaxy contamination. This is particularly significant in cases where the central aperture may not fully capture galaxy emissions (see \cref{subsec:4.3}).
    \item We expanded the X-ray to bolometric correction formula to include the low-luminosity regime, making it applicable across ten orders of magnitude in $L_\mathrm{Bol}$. While this formula shows lower $k_\mathrm{X}$ values for LLAGN compared to commonly assumed values, it remains consistent with high-luminosity AGN extrapolations from similar methodologies (see \cref{subsec:4.2}).
    \item Analysis of the $\alpha_\mathrm{ox}$ index revealed behavior that deviates from that observed in the high-luminosity regime. Unlike QSOs, where $\alpha_\mathrm{ox}$ correlates with $\lambda_\mathrm{Edd}$, LLAGN show a nearly constant or weakly correlated $\alpha_\mathrm{ox}$ across different accretion rates. This difference may suggest a shift in accretion physics and photon production mechanisms in LLAGN environments (see \cref{subsec:4.4} and \cref{fig:alpha_vs_lambda}).
\end{enumerate}

Our findings substantially advance the characterization and comprehension of LLAGN, a category of active galactic nuclei with significant analytical challenges. These insights pave the way for future explorations into LLAGN's distinctive attributes and impacts, emphasizing the critical role of a multi-wavelength perspective in AGN research.

\begin{acknowledgements}
We kindly thank Dra. Ramos-Almeida, Dr. Delvecchio, Dr. Fernández-Ontiveros and Dra. Pietro for valuable comments and discussion on this work. We also thank the anonymous referee and the SED fitting community for useful feedback that improved the quality of the paper. This action has received funding from the European Union’s Horizon 2020 research and innovation programme under Marie Skłodowska-Curie grant agreement  No 860744 "Big Data Applications for Black Hole Evolution Studies"  (BiD4BESt\footnote{\url{https://www.bid4best.org/}}) and the European Union’s Horizon 2020 research and innovation program under grant agreement no. 101004168, the XMM2ATHENA project. We acknowledge support from the grant ASI n. 2024-10-HH.0 “Attivit`a scientifiche per la missione Euclid – fase E”. IEL acknowledges support from the Cassini Fellowship program at INAF-OAS. EB acknowledges the support of the INAF Large Grant 2022 "The metal circle: a new sharp view of the baryon cycle up to Cosmic Dawn with the latest generation IFU facilities". SB acknowledges support from the Spanish Ministerio de Ciencia e Innovación through project PID2021-124243NB-C21. The color schemes used in this work are color-blind friendly from Paul Tol's Notes\footnote{\url{https://personal.sron.nl/~pault/}}. We also acknowledge the use of computational resources from the parallel computing cluster of the Open Physics Hub\footnote{\url{https://site.unibo.it/openphysicshub/en}} at the Physics and Astronomy Department of the University of Bologna.
\end{acknowledgements}

\bibliographystyle{aa}
\bibliography{manuscript.bib}

\begin{appendix} 
\section{Changes in Cigale inputs and outputs}
\label{appendix:cigale}

We present the input of the new \verb|lopez24| module in \cref{tab:in_phy}. Output physical parameters are presented in \cref{tab:out_phy}. Additionally, the table includes new quantities introduced by this work on SKIRTOR.

\begin{table*}
\centering
\setlength{\tabcolsep}{1.mm}
\caption{Input parameters for the new xray}
\label{tab:in_phy}
\begin{tabular}{lllcc} \hline\hline
Module & Parameters & Explanation & Default values & Units \\
\hline\\[-1.5ex]
    \multirow{5}{*}{\shortstack[l]{lopez24}} 
    & xray.E\_cut & Exponential cutoff energy of the AGN spectrum & 300 & keV \\
    & xray.alpha\_irx & The ratio between $\nu L_\nu (\mathrm{12~\mu m})$ and $L_\mathrm{X}(2$--$10~\mathrm{keV})$ & 0.0, 0.1, 0.2, 0.3, 0.4, 0.5, 0.6 & $-$ \\
    & xray.det\_hmxb & Deviation from the expected high-mass X-ray binary & 0. & $-$ \\
    & xray.det\_lmxb & Deviation from the expected low-mass X-ray binary & 0. & $-$ \\
    & xray.gam & Photon index ($\Gamma$) of the AGN intrinsic X-ray spectrum & 1.8 & $-$ \\ [0.5ex]
\hline
\end{tabular}
\end{table*}

\begin{table*}
\centering
\setlength{\tabcolsep}{1.mm}
\caption{Additional output physical parameters for the 
        new xray and SKIRTOR modules}
\label{tab:out_phy}
\begin{tabular}{llll} \hline\hline
Module & Parameters & Explanation & Units \\
\hline\\[-1.5ex]
    \multirow{6}{*}{\shortstack[l]{lopez24}} 
    & xray.agn\_Lnu\_2keV & The AGN $L_\nu$ at 2~keV & W Hz$^{-1}$ \\
    & xray.agn\_Lx\_2to10keV & The AGN 2--10~keV luminosity & W \\
    & xray.agn\_Lx\_total & The AGN total (0.25--1200~keV) X-ray luminosity & W \\
    & xray.lmxb\_Lx\_2to10keV & The 2--10~keV LMXB luminosity & W \\
    & xray.hmxb\_Lx\_2to10keV & The 2--10~keV HMXB luminosity & W \\
    & xray.hotgas\_Lx\_0p5to2keV & The 0.5--2~keV hot-gas luminosity & W \\[0.5ex]
\hline\\[-1.5ex]
    \multirow{9}{*}{\shortstack[l]{SKIRTOR}} 
    & agn.L\_12um & The total AGN disk luminosity at 12~$\mu m$ (sum of accretion and dust) & W \\ 
    & agn.L\_6um & The total AGN disk luminosity at 6~$\mu m$ (sum of accretion and dust) & W \\ 
    & agn.accretion\_power & The intrinsic AGN disk luminosity averaged over all directions & W \\[0.5ex]
    & agn.disk\_luminosity & The observed AGN disk luminosity (might be extincted) & W \\ 
    & agn.intrin\_Lnu\_2500A & The intrinsic AGN $L_\nu$ at 2500~\AA\ at
                               viewing angle $=30^\circ$ & W Hz$^{-1}$ \\
    & agn.luminosity & The sum of agn.disk\_luminosity and agn.total\_dust\_luminosity & W \\
    & agn.polar\_dust\_luminosity & The AGN polar dust reemitted luminosity & W \\
    & agn.torus\_dust\_luminosity & The AGN toroid dust reemitted luminosity& W \\
    & agn.total\_dust\_luminosity & The AGN total dust reemitted luminosity & W \\
\hline
\end{tabular}
\end{table*}

\section{Behavior in smaller apertures}
\label{appendix:aperture}

As discussed in \cref{subsec:3.1}, the aperture size in this study was selected to encompass the larger PSFs of the FIR bands.  This choice allows us to constrain MIR emission from both the host galaxy and the AGN, thus optimizing the application of the $L_\mathrm{X}$--$L_\mathrm{12\mu m}$ relation. Our mock analysis demonstrated that a higher value of agn.frac improves constraints on AGN parameters. By using a smaller aperture, we could reduce host galaxy contamination, thereby increasing agn.frac and providing a clearer view of the AGN itself.

\begin{figure}[]
    \centering
    \includegraphics[width=0.99\linewidth]{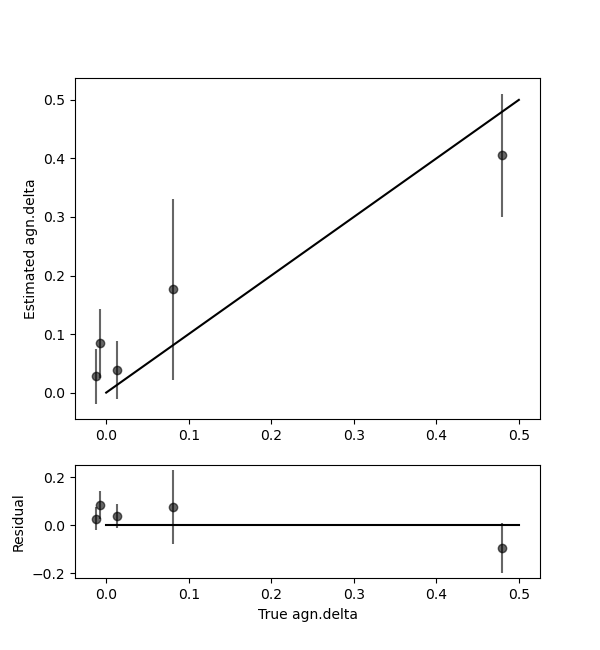}
    \caption[]{Similar to \cref{fig:parameters}, this figure shows the comparison between the true $\delta_\mathrm{AGN}$ (from the best-fit SED) and the estimated values from the re-performed SED fitting. A reduced aperture improves the constraints on this parameter.}
    \label{fig:small}
\end{figure}

However, selecting a smaller aperture that excludes MIR and FIR emission is problematic, as it would limit our ability to apply the $L_\mathrm{X}$--$L_\mathrm{12\mu m}$ prior. To explore the impact of a smaller aperture, we reanalyzed five sources that are also included in \citet{FernandezOntiveros23}, using their high-resolution sub-arcsecond photometry in conjunction with our module. With this sub-arcsecond data, and following the procedure outlined in \cref{subsec:4.1}, the AGN delta parameters showed significantly improved constraints (see \cref{fig:small}).

While these preliminary results are promising, the small number of sources limits our ability to generalize these findings to the broader LLAGN population. Future observations from \textit{JWST} and \textit{Euclid}, with their higher resolution in the near and mid-IR could provide better AGN emission estimation, reducing host galaxy contamination and improving constraints on AGN parameters. Next-generation FIR instruments with enhanced angular resolution could further aid in constraining FIR emission, thereby refining parameter estimates across a larger sample.

\end{appendix}

\end{document}